\documentclass[12pt]{article}
\usepackage{graphicx,psfrag,layout,color,amssymb,amsfonts,amsmath,psfig,times,epsf,epsfig,boxedminipage,url,xspace}
\def\ket#1{|#1\rangle}

\begin{document}
\title{ A study on the shape of two-photon wavefunctions after the nonlinear interaction with \\ a one-dimensional atom}
\author{Kunihiro Kojima$^{1}$\footnote{kuni@es.hokudai.ac.jp}, Holger F. Hofmann$^{1,2}$\footnote{hofmann@es.hokudai.ac.jp}, Shigeki Takeuchi$^{1,2}$\footnote{takeuchi@es.hokudai.ac.jp, tel +81-11-706-2646, fax +81-11-706-4956} and Keiji Sasaki$^{1}$\footnote{sasaki@es.hokudai.ac.jp}\\ {\small $^1$ \it Research Institute for Electronic Science, Hokkaido University,}\\ {\small \it Kita-12, Sapporo 060-0812, Japan}\\ {\small $^2$ \it PRESTO, Japan Science and Technology Corporation (JST),}\\ {\small \it Hokkaido University, Kita-12 Nishi-6, Kita-ku, Sapporo 060-0812, Japan}}
\maketitle
\begin{abstract}
We study the interaction of Gaussian one- and two-photon pulses with a single two-level atom based on a one-dimensional model of pulse propagation to and from the atom. The characteristic time scale of the atomic response is the dipole relaxation time $1/\Gamma$. We therefore compare the effect of the non-linear two-photon interaction for a long pulse length of $10/\Gamma$ with a short pulse of $1/\Gamma$. Our results indicate that the effect of the non-linear interaction is particularly strong for the short pulse length of $1/\Gamma$.\\
\\
\vspace{0.9cm}
{{\it Keywords:} Full quantum theory of light-atom interaction, Cavity QED, Non-linear photon-\\[-0.9cm]photon interaction, two-photon wavefunction, Single atom nonlinearity, One-dimensional atom.}
\end{abstract}
\section{\label{sec:level1}INTRODUCTION}
Nonlinearities are important in optical systems since they enable us to manipulate light by light. Especially, nonlinearities sensitive to single photons are necessary for quantum optical applications such as optical quantum computation \cite{nielsen0}, entanglement generation \cite{holger3}, and quantum non-demolition measurements \cite{nondemo}. In particular, strong nonlinearities are indispensable for quantum computation, since conditional operations at the single photon level are required to realize elementary gate operations. However, conventional optical media require very high photon densities to obtain significant nonlinear effects \cite{mills}. Single atom cavity quantumelectrodynamics may offer a possible solution \cite{bermann,turchette}. In particular, a cavity can focus all the light in an input beam on a single atom, effectively creating a situation where a single atom is coupled to a  one-dimensional light field \cite{holger2}. Such a one-dimensional atom is an ideal device for the non-linear manipulation of few photon pulses. It is therefore of great interest to investigate the response of a one-dimensional atom to one- and two-photon input pulses.

There are many theories for such atom-cavity systems \cite{bermann}. However these theories usually eliminate the quantum state of the field outside the atom-cavity system and therefore the analysis of the precise spatiotemporal coherence of the input and output photon pulses can not be achieved. In our previous work, we have therefore presented a theory for one- and two-photon pulses that includes the propagation to and from the system in a bad-cavity regime \cite{kojima,holger3}, based on a one-dimensional model of light-atom interaction \cite{holger0}. In these studies, we focused on the changes of the photon-photon correlations in time \cite{kojima} and in frequency \cite{holger3}.

In this paper, we study the effect of the non-linear interaction with the one-dimensional atom on the shape of one- and two-photon wavefunctions using Gaussian input pulses of different pulse lengths. For this purpose, we first introduce an analysis of the interaction in terms of absorption and transmission processes in section \ref{sec:level2}. The characteristic time scale defining the response of the atom is given by the dipole relaxation time $1/\Gamma$. In section \ref{sec:level3}, we then analyze the interaction processes for a long pulse length of $10/\Gamma$ in terms of pulse amplitude and pulse delay time. From these results, we can estimate that a maximal non-linear effect should be obtained for a pulse length equal to the dipole relaxation time of the atom. In section \ref{sec:level4}, we therefore analyze the corresponding interaction processes for a short pulse length of $1/\Gamma$, confirming the dominance of the non-linear effect throughout the short two-photon pulse. The results of the analysis are summarized and discussed in section \ref{sec:level5}. 
\section{\label{sec:level2}MODEL AND THEORY}
\subsection{Theoretical model and physical realization}
\begin{figure}[htbp]
\vspace{0.3cm}
\begin{picture}(0,0)
\put(-10,160){(a)}
\put(270,160){(b)}
\end{picture}
\includegraphics[width=9cm]{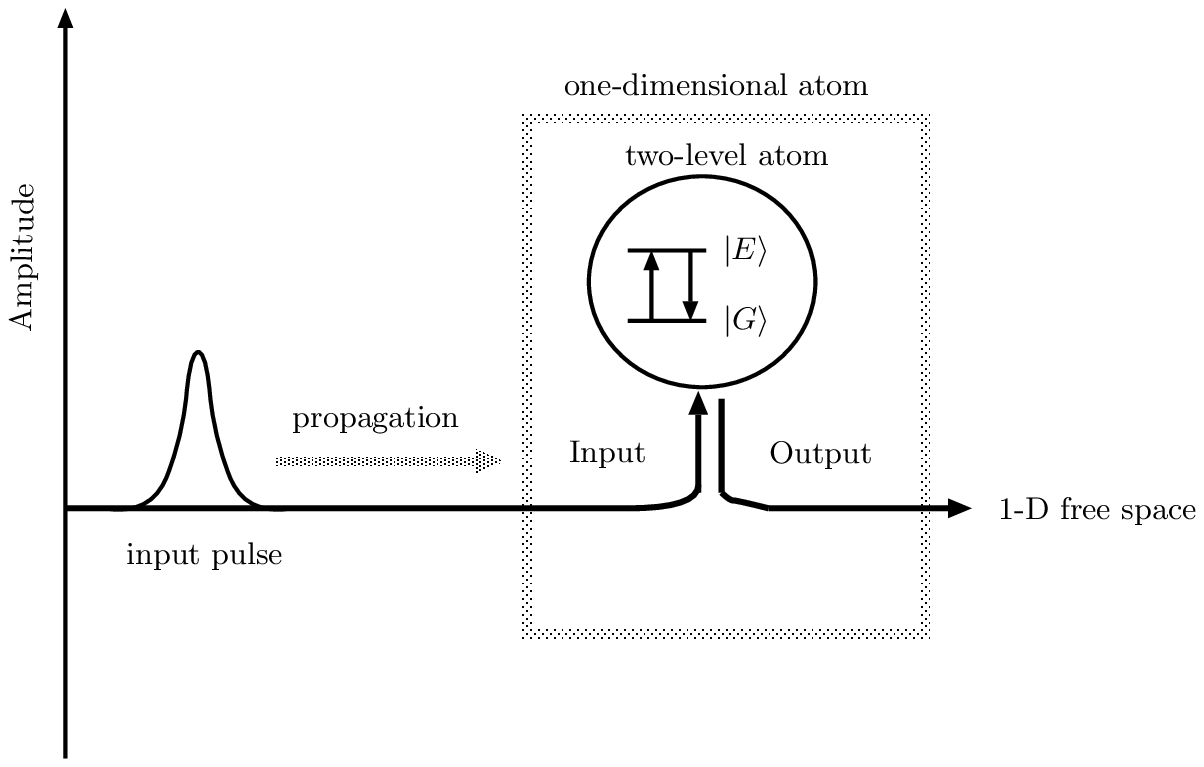}
\ \ \ \ \ \ \ \ \ \ \ 
\includegraphics[width=6.5cm]{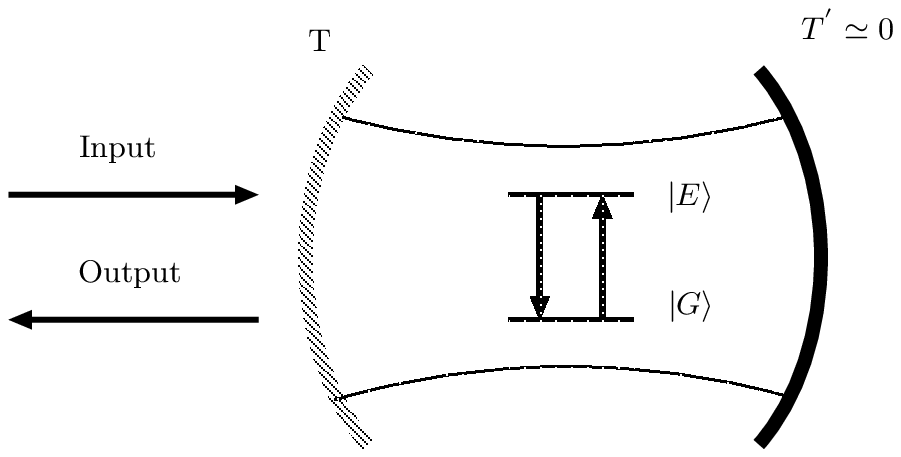}
\vspace{0.3cm}
\caption{\label{fig:figure1}(a) Schematic of a two-level atom in one-dimensional free space and (b) physical realization of the one-dimensional atom using an atom-cavity system. An incoming one-photon or two-photon wavepacket interacts with the atom locally, and the output wavepacket is then emitted into the output field. The input field of the one-dimensional free space shown in (a) corresponds to the input of the one-sided cavity in (b). Likewise, the output field of the one-dimensional free space shown in (a) corresponds to the output of the one-sided cavity in (b).}
\end{figure}

The model of a single two-level atom in one-dimensional free space is shown in figure.~\ref{fig:figure1}(a). In this model, an incoming one-photon or two-photon wavepacket interacts with the atom locally, and the output wavepacket is then emitted into the output field. The physical realization of this model can be implemented by a two-level atom coupled with a single mode of a one-sided cavity in the bad cavity regime \cite{holger2,kojima}. The cavity geometry is shown in figure.~\ref{fig:figure1} (b). The input field of the one-dimensional free space shown in figure.~\ref{fig:figure1} (a) corresponds to the input of the one-sided cavity in figure.~\ref{fig:figure1} (b). Likewise, the output field in figure.~\ref{fig:figure1} (a) corresponds to the output of the one-sided cavity in figure.~\ref{fig:figure1} (b). Here, we assume that the spontaneous emission rate through the cavity mode is much larger than the spontaneous emission rate through non-cavity modes. In terms of the conventional cavity quantum electrodynamics parameters, this regime is characterized by \(\kappa \gg g \gg \gamma\), where \(\kappa\) is the cavity damping rate through the left mirror, \(g\) is the dipole coupling between the atom and the cavity mode and $\gamma$ is the rate of spontaneous emission into the non-cavity modes. Since the cavity damping rate \(\kappa\) is much faster than the dipole coupling \(g\), the method of adiabatic elimination can be applied to the time evolution of the cavity field \cite{rice}. This means that the interaction between the atom and the outside field through the cavity field can be expressed by an effective dipole relaxation rate \(\Gamma = g^{2}/\kappa\). The dipole relaxation rate $\Gamma$ describes the dipole damping caused by emissions through the left mirror of the cavity, and the corresponding rate of spontaneous emission through the cavity is equal to \(2\Gamma\) \cite{rice,turchetteb,holger2}. In our case, we assume that the rate of spontaneous emission into the non-cavity modes \(\gamma\) is negligible (\(\gamma \ll 2\Gamma\)). Nearly all emissions from the atom can then be confined to the cavity and 2\(\Gamma\) is the total spontaneous emission rate of the excited atom in the cavity. In present cavity designs, this can be realized by exploiting the Purcell effect. For example, in the case of Turchette et al.'s experiment \cite{turchette}, the cavity parameters indicate that about 80\% of the sponaneous emission from the atom is emitted through the single cavity mode.
\subsection{General solution of the interaction dynamics}
In the following, we investigate the effect of the one-dimensional atom on one and two-photon input pulses of different pulse length.  For this purpose, we have to formulate the quantum states for a given input pulse shape for both one photon and two photon cases. If the real space representation for the one-photon state reads,
\begin{eqnarray}
\ket{\text{A}}=\int \ dx \Psi_{\text{A}}(x) \ket{x}, \label{eq:onephotonsate}
\end{eqnarray}
the two-photon state can be expressed as the product state
\begin{eqnarray}
\ket{\psi}_{\text{in}} &=& \ket{\text{A}_{1}} \otimes \ket{\text{A}_{2}} = \int \ dx_{1}dx_{2} \Psi_{\text{A}}(x_{1}) \cdot \Psi_{\text{A}}(x_{2}) \ket{x_{1};x_{2}}.\label{eq:spatialinput}
\end{eqnarray}
The light pulses characterized by the above input states are propagating at the velocity of light in the input field of the schematic shown in figure. \ref{fig:figure1}(a). To simplify the expression, the spatial coordinates, $x_{1}$ and $x_{2}$, therefore represent a coordinate system moving at the velocity of light. The indices 1 and 2 of $x_{1}$ and $x_{2}$ in eq.~(\ref{eq:spatialinput}) identify the two particles. The spatial features of the probability amplitudes given by the wavefunction $\Psi_{\text{A}}$ correspond to the spatial features of the pulse mode A. Thus, the wavefunction $\Psi_{\text{A}}(x_{1}) \cdot \Psi_{\text{A}}(x_{2})$ characterizes the single mode two-photon wavefunction, where the wavefunctions of both particles overlap perfectly.
%2.3 Response function
In our previous work \cite{kojima,holger3}, we derived the unitary time evolution operator in the Hilbert spaces for the one photon component and the two photon component by solving the Schr\"odinger equation of the interaction between a one-dimensional field and a single atom. The output one-photon state in the far field can then be obtained by convoluting the one-photon input wavefunction $\Psi_{\text{A}}(x)$ with the matrix description of the unitary operator in real space,
\begin{eqnarray}
&& \ket{\psi}_{\text{1photon}} = \int \ dx\ \Psi_{\text{out}}(x) \ket{x}  \nonumber
\end{eqnarray}
\begin{eqnarray}
\text{with} && \Psi_{\text{out}}(x) = \int dx^{'} \ {\bf u}_{\text{1photon}} (x;x^{'}) \cdot \Psi_{\text{A}}(x^{'}). \label{eq:oneout}
\end{eqnarray}

The matrix elements ${\bf u}_{\text{1photon}}(x,x^{'})$ are given by
\begin{eqnarray} 
&& {\bf u}_{\text{1photon}}(x;x^{'}) = \delta(x-x^{'})-\frac{2\Gamma}{c} e^{-\frac{\Gamma}{c}(x^{'}-x)} \ \ \ \text{ for $x \leq x^{'}$, else $0$.}\label{eq:onemat}
\end{eqnarray}
Likewise, the output two-photon state in the far field can be obtained by a linear convolution, 
\begin{eqnarray}
&& \ket{\psi}_{\text{out}} = \int \ dx_{1}dx_{2} \Psi_{\text{out}}(x_{1};x_{2}) \ket{x_{1};x_{2}},  \nonumber
\end{eqnarray}
\begin{eqnarray}
\text{with} && \Psi_{\text{out}}(x_{1};x_{2}) = \int dx^{'}_{1}dx^{'}_{2} \ {\bf u} (x_{1},x_{2};x^{'}_{1},x^{'}_{2}) \cdot \Psi_{\text{A}}(x^{'}_{1}) \cdot \Psi_{\text{A}}(x^{'}_{2}). \label{eq:actual}
\end{eqnarray}
The matrix elements ${\bf u} (x_{1},x_{2};x^{'}_{1},x^{'}_{2})$ are given by
\begin{eqnarray} 
&& {\bf u}(x_{1},x_{2};x_{1}^{'},x_{2}^{'}) = {\bf u}_{\text{1photon}}(x_{1};x_{1}^{'}) \cdot {\bf u}_{\text{1photon}}(x_{2};x_{2}^{'}) + \Delta {\bf u}^{{\bf Nonlin}}(x_{1},x_{2};x_{1}^{'},x_{2}^{'}), \nonumber
\end{eqnarray}
where the contribution of the non-linear photon-photon interaction is given by
\begin{eqnarray}
\Delta {\bf u}^{{\bf Nonlin}}(x_{1},x_{2};x_{1}^{'},x_{2}^{'}) = -\frac{4 \Gamma^{2}}{c^{2}} e^{-\frac{\Gamma}{c}(x_{1}^{'}+x_{2}^{'}-x_{1}-x_{2})}\ \ \ \text{ for }x_{1},x_{2} < {\bf Min}[x_{1}^{'},x_{2}^{'}]. \label{eq:nonlineffect}
\end{eqnarray}
The output wavefunctions $\Psi_{\text{out}}(x)$ and $\Psi_{\text{out}}(x_{1};x_{2})$ describe the spatial feature of the output states in the far field of the atom shown in figure.~\ref{fig:figure1} (a). Note that the spatiotemporal features of the input one-photon and two-photon wavefunctions are generally not preserved in the output due to the atomic response described by the matrix elements ${\bf u}_{\text{1photon}}(x;x^{'})$ and $\Delta {\bf u}^{{\bf Nonlin}}(x_{1},x_{2};x_{1}^{'},x_{2}^{'})$.

\subsection{Components of the interaction}
The matrix element of the time evolution \({\bf u}_{\text{1photon}}(x;x^{'})\) in eq.(\ref{eq:onemat}) can be expanded in terms of two interaction processes,
\begin{eqnarray}
{\bf u}_{\text{1photon}}(x;x^{'}) = {\bf u}_{\text{prop}}(x;x^{'}) + {\bf u}_{\text{abs}}(x;x^{'}),
\end{eqnarray}
where the two processes are (I) single photon transmission without absorption, given by
\begin{eqnarray}
{\bf u}_{\text{prop}}(x;x^{'}) = \delta(x-x^{'}),
\end{eqnarray}
and (II) single photon reemission after absorption, given by
\begin{eqnarray}
{\bf u}_{\text{abs}}(x;x^{'}) = -\frac{2\Gamma}{c} e^{-\frac{\Gamma}{c}(x^{'}-x)} \ \ \ \text{ for $x \leq x^{'}$, else $0$.} \label{eq:absgenerater}
\end{eqnarray}
The output one-photon wavefunction $\Psi_{\text{out}}(x)$ can thus be interpreted as the result of quantum interference of the two processes,
\begin{eqnarray}
\Psi_{\text{out}}(x) &=& \Psi_{\text{prop}}(x) + \Psi_{\text{abs}}(x),
\end{eqnarray}
where the components are given by 
\begin{eqnarray}
\Psi_{\text{\text{prop}}}(x)  &=& \int^{\infty}_{-\infty} dx^{'}{\bf u}_{\text{prop}}(x,x^{'}) \cdot
 \Psi_{\text{A}}(x^{'}) = \Psi_{\text{A}}(x) \nonumber\\
\Psi_{\text{\text{abs}}}(x)  &=& \int^{\infty}_{-\infty} dx^{'}{\bf u}_{\text{abs}}(x,x^{'}) \cdot
 \Psi_{\text{A}}(x^{'}). \label{eq:oneoutputcomponents}
\end{eqnarray}

In the same way, the matrix element of the time evolution \({\bf u}(x_{1},x_{2};x_{1}^{'},x_{2}^{'})\) can be expanded in terms of three interaction processes. Process (1) is the transmission of both photons without absorption. Process (2) is the transmission of one photon without absorption and the absorption and reemission of the other photon. Process (3) is the absorption and reemission of both photons. In the following, we will refer to process (1) as the transmission process, to process (2) as the one photon absorption process, and to process (3) as the two photon absorption process. The transformation of the two-photon wavefunction can then be written as
\begin{eqnarray}
{\bf u}(x_{1},x_{2};x_{1}^{'},x_{2}^{'}) &=& {\bf u}^{\text{(1)}}(x_{1},x_{2};x_{1}^{'},x_{2}^{'}) + {\bf u}^{\text{(2)}}(x_{1},x_{2};x_{1}^{'},x_{2}^{'}) + {\bf u}^{\text{(3)}}(x_{1},x_{2};x_{1}^{'},x_{2}^{'}).
\end{eqnarray}
The components of \({\bf u}(x_{1},x_{2};x_{1}^{'},x_{2}^{'})\) read
\begin{eqnarray}
{\bf u}^{\text{(1)}}(x_{1},x_{2};x_{1}^{'},x_{2}^{'}) &=& {\bf u}_{\text{prop}}(x_{1};x^{'}_{1}) \cdot {\bf u}_{\text{prop}}(x_{2};x^{'}_{2}) \nonumber\\
{\bf u}^{\text{(2)}}(x_{1},x_{2};x_{1}^{'},x_{2}^{'}) &=& {\bf u}_{\text{prop}}(x_{1};x^{'}_{1}) \cdot {\bf u}_{\text{abs}}(x_{2};x^{'}_{2})  +  {\bf u}_{\text{abs}}(x_{1};x^{'}_{1}) \cdot {\bf u}_{\text{prop}}(x_{2};x^{'}_{2})\nonumber\\
{\bf u}^{\text{(3)}}(x_{1},x_{2};x_{1}^{'},x_{2}^{'}) &=& {\bf u}_{\text{abs}}(x_{1};x^{'}_{1}) \cdot {\bf u}_{\text{abs}}(x_{2};x^{'}_{2}) + \Delta {\bf u}^{{\bf Nonlin}}(x_{1},x_{2};x_{1}^{'},x_{2}^{'}). \label{eq:twophotonprocesses}
\end{eqnarray}
Note that the non-linear contribution $ \Delta {\bf u}^{{\bf Nonlin}}$ only occurs in the two photon absorption process. This is because the non-linearity is due to the saturation of the two-level atom, which prevents the absorption of two photons at the same time.

Likewise, the output wavefunction \(\Psi_{\text{out}}(x_{1},x_{2})\) can be expanded as
\begin{eqnarray}
\Psi_{\text{out}}(x_{1},x_{2}) &=& \Psi^{\text{(1)}}_{\text{out}}(x_{1},x_{2}) + \Psi^{\text{(2)}}_{\text{out}}(x_{1},x_{2}) + \Psi^{\text{(3)}}_{\text{out}}(x_{1},x_{2}),
\end{eqnarray}
where the $\Psi^{\text{({\it i})}}(x_{1},x_{2})$ are given by 
\begin{eqnarray}
\Psi_{\text{out}}^{\text{(i)}}(x_{1},x_{2})  &=&
 \int^{\infty}_{-\infty} dx^{'}_{1}dx^{'}_{2} {\bf u}^{\text{({\it i})}}(x_{1},x_{2},x_{1}^{'},x_{2}^{'}) \cdot
 \Psi_{\text{in}}(x^{'}_{1},x^{'}_{2})\ \ \ \ \ \ \ \ \ \ \text{({\it i}=1,2,3)}. \label{eq:outputcomponents}
\end{eqnarray}
This analysis will be convenient to understand the origin of the statistical properties of the output photons. In the next section, these results are applied to the cases of Gaussian one-photon and two-photon input wavepackets.
\subsection{Definition of input pulses}
The dependence of the coherent spatiotemporal correlations between the output photons on the input pulse length is analyzed numerically using the results in the previous subsection. For our analysis, we use Gaussian one-photon and two-photon input wavefunctions,
\begin{eqnarray}
\Psi_{\text{A}}(x) &=& e^{-\left(x^{2}\right)/(cT)^{2}}/\sqrt{N} \nonumber\\
\text{with} && N = \sqrt{(\pi/2)(cT)^{2}}.
\end{eqnarray} 
The pulse length $T$ is defined as twice the dispersion of the Gaussian distribution of the photons in time. Note that the response of the atom is governed by the time scale $1/\Gamma$, therefore the atom responds differently to long pulses and short pulses. In the following, the term 'long pulse' refers to a pulse whose pulse length $T$ is much longer than the dipole relaxation time $1/\Gamma$. Likewise, the term 'short pulse' refers to a pulse whose pulse length $T$ is about equal to the dipole relaxation time $1/\Gamma$. In the next section, we analyze the output wavefunctions for long one-photon and two-photon pulses in order to identify the characteristic quantum interference effects.    
\section{\label{sec:level3} RESPONSE TO A LONG INPUT PULSE}
\subsection{Explanation of the one-photon output wavepacket in terms of one photon processes}
\begin{figure}[htbp]
\psfrag{x}[t][t]{$x\ (\text{in units of}\ c/\Gamma)$}
\psfrag{amp}[b][b]{$\Psi_{\text{in/out}}(x)\ (\text{in units of}\ \Gamma/c)$}
\psfrag{amp1}[b][b]{$\Psi^{\text{(I)},\ \text{(II)}}_{\text{out}}(x)\ (\text{in units of}\ \Gamma/c)$}
\begin{minipage}{.45\linewidth}
\begin{picture}(0,0)
\put(0,200){(a)}
\end{picture}
\includegraphics[width=\linewidth,height=\linewidth]{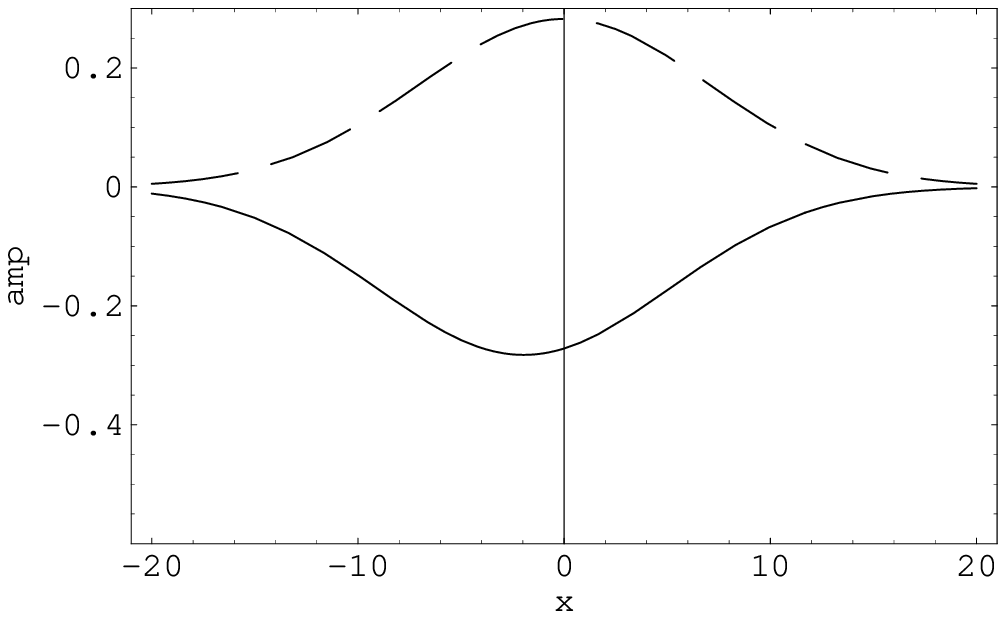}
\end{minipage}
\ \ \ \ \ \ \ \ \ \ \ \ 
\begin{minipage}{.45\linewidth}
\begin{picture}(0,0)
\put(0,200){(b)}
\end{picture}
\includegraphics[width=\linewidth,height=\linewidth]{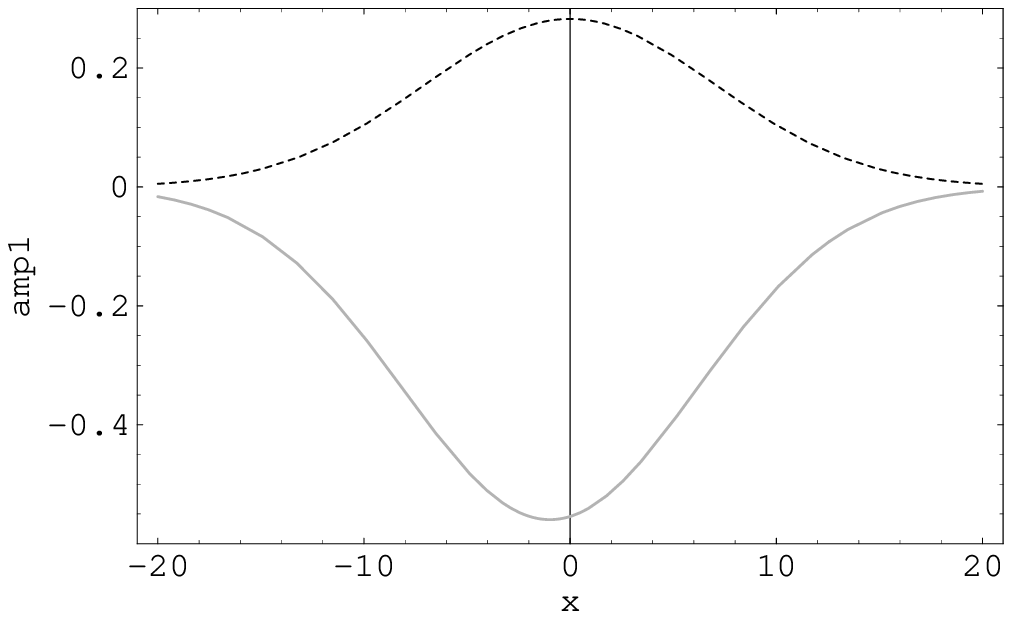}
\end{minipage}
\caption{\label{fig:figure2}Gaussian input and output one-photon wave packets for an input wave packet length of \(T=10/\Gamma\). The horizontal axes \(x\) represents the spatial coordinate of one photon. In (a), the broken line corresponds to the input wavefunction $\Psi_{\text{A}}(x)$ and the solid line corresponds to the output wavefunction $\Psi_{\text{out}}(x)$. In (b), the dotted line corresponds to the component of the process (I): single photon transmission without absorption and the thin line corresponds to the component of the process (II): single photon reemission after absorption.}
\end{figure}

Figure.~\ref{fig:figure2} (a) shows an example of a one-photon long pulse (broken line) and the corresponding output wavefunction (solid line). In this example, we have chosen an input pulse length of \(T=10/\Gamma\) which is 10 times greater than the dipole relaxation time \(1/\Gamma\).

The pulse shape of the output one-photon pulse is almost identical with the input pulse except for the $\pi$ phase flip in the output amplitudes compared to the input amplitudes, $\Psi_{\text{out}}(x) \approx -\Psi_{\text{A}}(x)$.  Furthermore the minimum of the output amplitude is delayed compared to the maximum of the input amplitudes. The delay time is about $2/\Gamma$. Let us now analyze the origin of these features.

Figure.~\ref{fig:figure2} (b) shows the components corresponding to different light-atom interaction processes as given by eqs.(\ref{eq:oneoutputcomponents}). The dotted line is the component of single photon transmission without absorption, $\Psi_{\text{prop}}(x)=\Psi^{\text{A}}(x)$ (see eqs.(\ref{eq:oneoutputcomponents})). The thin line is the component of single photon absorption, $\Psi_{\text{abs}}(x)$. Since the absorption process convolutes the input amplitude with an exponential function according to eqs.~(\ref{eq:oneoutputcomponents}), this component is not a precise Gaussian anymore. However, in the long pulse limit, the convolution function is so much narrower than the Gaussian that the change in pulse shape is negligibly small. The component $\Psi_{\text{abs}}(x)$ can then be approximated as
\begin{eqnarray}
\Psi_{\text{abs}}(x) &\approx&  -2 \Psi_{\text{A}}(x) \label{eq:appabs}
\end{eqnarray}
As this approximation indicates, the amplitude of the absorption process is about two times as large as the amplitude of the transmission process. It is then possible to understand the $\pi$ phase flip in the output as a result of interference, 
\begin{eqnarray}
\Psi_{\text{out}}(x) &\approx& \underbrace{\Psi_{\text{A}}(x)}_{\text{transmission}}\underbrace{-\ 2\ \Psi_{\text{A}}(x)}_{\text{absorption}}=\ -\Psi_{\text{A}}(x), \label{eq:apponeoutput}
\end{eqnarray}
where the negative amplitude of the absorption process is twice as high as the positive amplitude of the transmission process. 

However this approximation cannot explain the delay of the peak of the output wavefunction in figure.~2 (a). In the long pulse limit, the delay time of $1/\Gamma$ in $\Psi_{\text{abs}}(x)$ can be included in the approximation given in eq.(\ref{eq:appabs}) by expanding $\Psi_{\text{abs}}(x)$ as follows,  
\begin{eqnarray}
\Psi_{\text{abs}}(x) &\approx& -2 \Psi_{\text{A}}(x+c/\Gamma) \nonumber \\
                     &\approx&  -2 \Psi_{\text{A}}(x)-2c/\Gamma \frac{\partial}{\partial x}\Psi_{\text{A}}(x). \label{eq:appabs2}
\end{eqnarray}
The delay time of $\Psi_{\text{out}}(x)$ can then be obtained from the interference of $\Psi_{\text{prop}}(x)$ and $\Psi_{\text{abs}}(x)$,
\begin{eqnarray}
\Psi_{\text{out}}(x) &\approx& \Psi_{\text{A}}(x) + \left(-2\Psi_{\text{A}}(x)-2c/\Gamma \frac{\partial}{\partial x}\Psi_{\text{A}}(x)\right) \nonumber\\
                     &=& -\left(\Psi_{\text{A}}(x)+2c/\Gamma \frac{\partial}{\partial x}\Psi_{\text{A}}(x)\right) \nonumber\\
                     &\approx& -\Psi_{\text{A}}(x+2c/\Gamma).
\end{eqnarray}   
The delay time of the output wavefunction is therefore equal to $2/\Gamma$, as seen in figure.~\ref{fig:figure2} (a).

\subsection{Explanation of the two-photon output wavepacket in terms of two photon processes}
Let us now consider the output wavefunction for the long two photon pulse. The output wavefunction can be analyzed in the same way as the one photon case, using the three interaction processes discussed in section \ref{sec:level2}.3. However, there are now two photon coordinates, $x_{1}$ and $x_{2}$, and, as shall be seen in the following, the distance $\left|x_{1}-x_{2}\right|$ between the two photons significantly influences the two photon absorption process. Note also that the bosonic symmetry of the indistinguishable photons requires that $\Psi(x_{1},x_{2})=\Psi(x_{2},x_{1})$.
 
\begin{figure}[htbp]
\psfrag{x1}[t][t]{$x_{1}\ (\text{in units of}\ c/\Gamma)$}
\psfrag{x2}[b][b]{$x_{2}\ (\text{in units of}\ c/\Gamma)$}
\psfrag{crossx}[t][t]{$\left(x_{1}+x_{2}\right)/2\ \ (\text{in units of}\ c/\Gamma)$}
\psfrag{crossamp1}[][]{$\Psi_{\text{in}}(x_{1},x_{2}=x_{1}+\tau)\ (\text{in units of}\ \Gamma/c)$}
\begin{minipage}{.45\linewidth}
\psfrag{20}[tr][tr]{\footnotesize 20}
\psfrag{10}[tr][tr]{\footnotesize 10}
\psfrag{0}[tr][tr]{\footnotesize 0}
\psfrag{-10}[tr][tr]{\footnotesize -10}
\psfrag{-20}[tr][tr]{\footnotesize -20}
\begin{picture}(0,0)
\put(-5,200){(a)}
\end{picture}
\includegraphics[width=\linewidth,height=\linewidth]{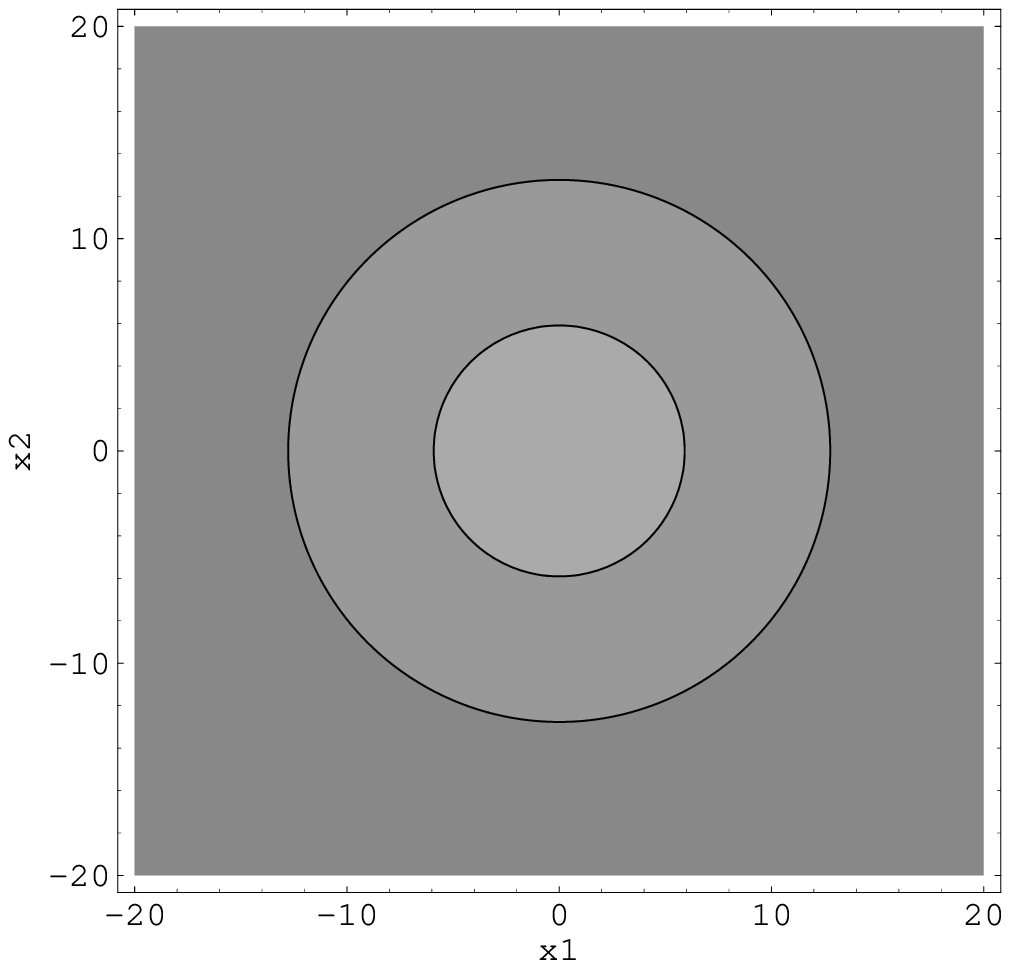}
\end{minipage}
\ \ \ \ \ \ \ \ \ \ \ \
\vspace{1cm} 
\begin{minipage}{.45\linewidth}
\begin{picture}(0,0)
\put(-20,200){(b)}
\end{picture}
\includegraphics[width=\linewidth,height=\linewidth]{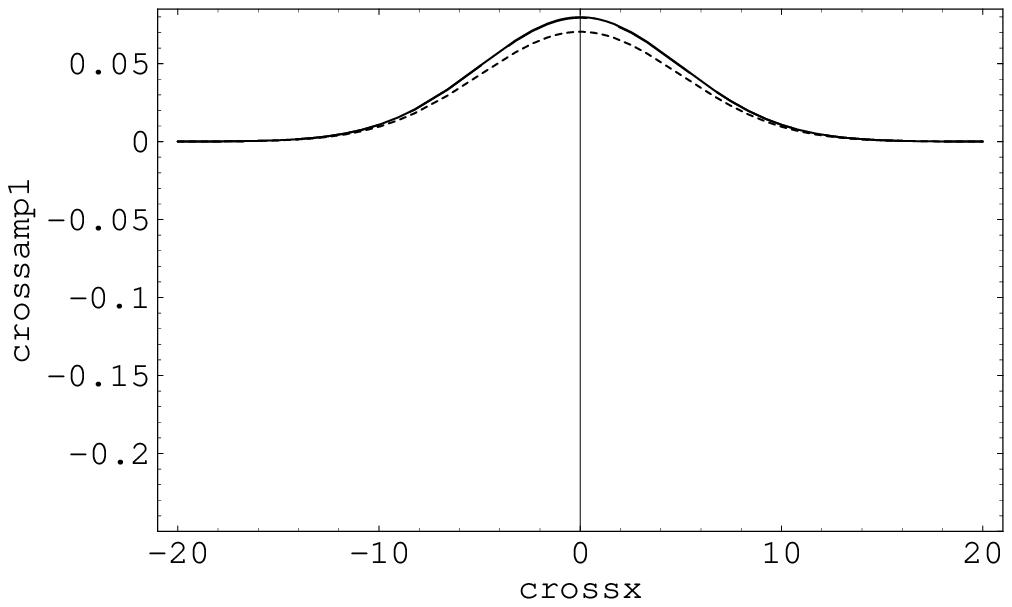}
\end{minipage}
\psfrag{space}[t][t]{$\left(x_{1}+x_{2}\right)/2\ \ (\text{in units of}\ c/\Gamma)$}
\psfrag{crossamp2}[][]{$\Psi_{\text{out}}(x_{1},x_{2}=x_{1}+c\tau)\ (\text{in units of}\ \Gamma/c)$}
\vspace{1cm}
\begin{minipage}{.45\linewidth}
\psfrag{20}[tr][tr]{\footnotesize 20}
\psfrag{10}[tr][tr]{\footnotesize 10}
\psfrag{0}[tr][tr]{\footnotesize 0}
\psfrag{-10}[tr][tr]{\footnotesize -10}
\psfrag{-20}[tr][tr]{\footnotesize -20}
\begin{picture}(0,0)
\put(-5,200){(c)}
\end{picture}
\includegraphics[width=\linewidth,height=\linewidth]{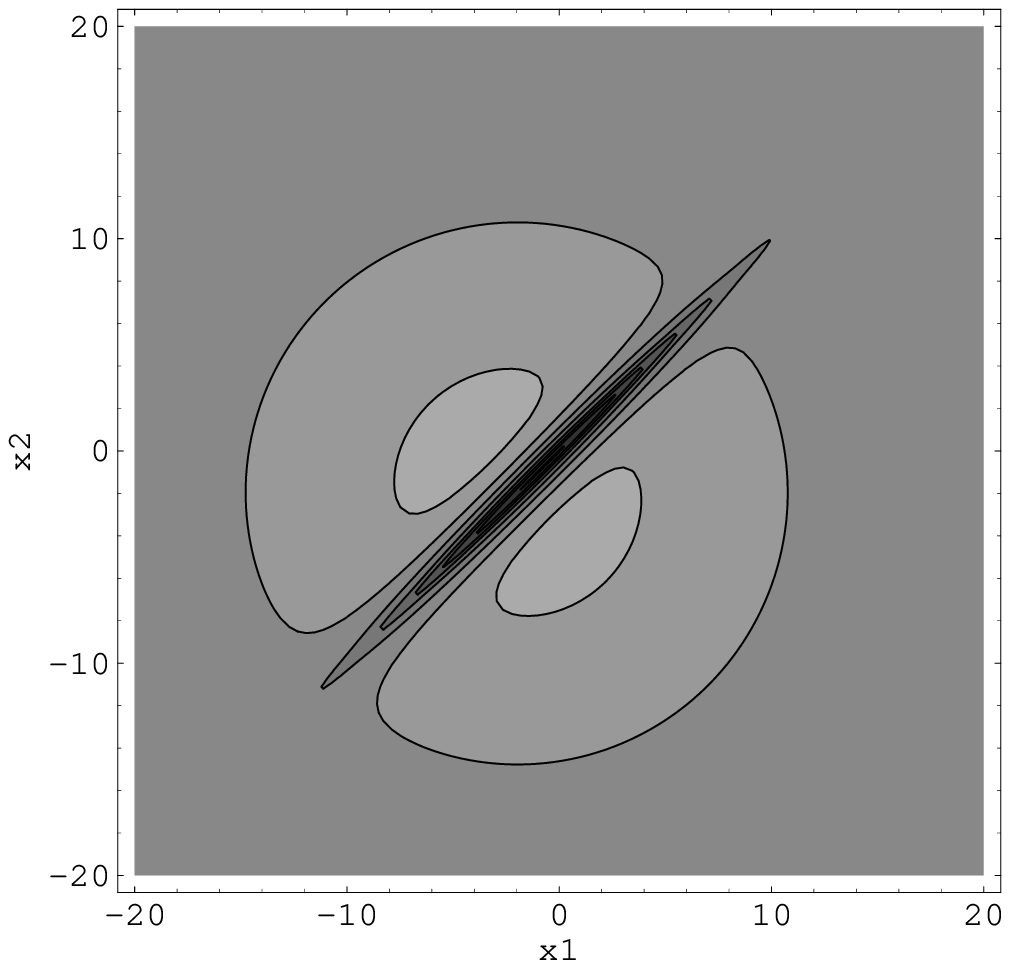}
\end{minipage}
\ \ \ \ \ \ \ \ \ \ \ \ 
\begin{minipage}{.45\linewidth}
\begin{picture}(0,0)
\put(-20,200){(d)}
\end{picture}
\includegraphics[width=\linewidth,height=\linewidth]{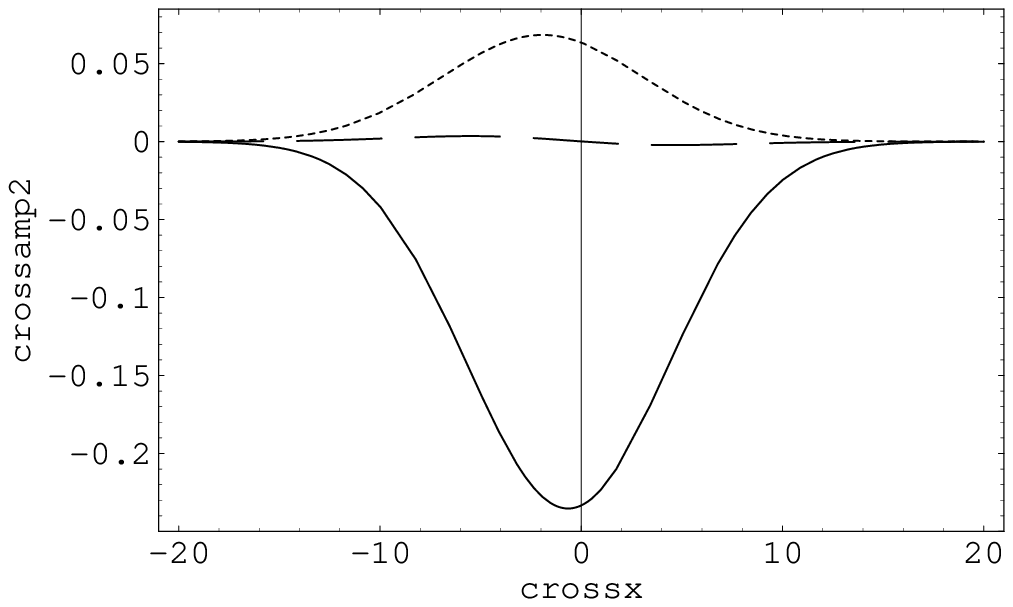}
\end{minipage}
\begin{minipage}{.45\linewidth}
\begin{picture}(0,0)
\put(-5,35){(e)}
\put(-5,-15){-0.35}
\put(190,-15){0.3}
\put(110,-15){0}
\end{picture}
\includegraphics[width=\linewidth,height=1cm]{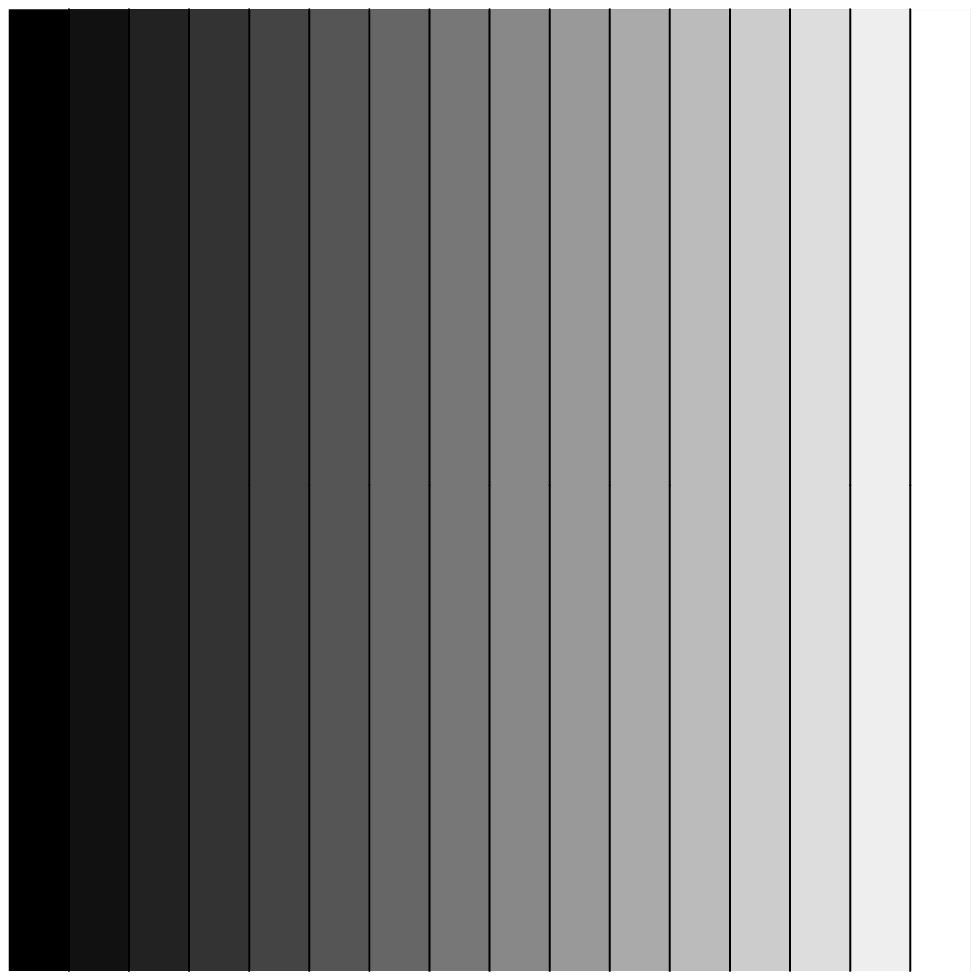}
\end{minipage}
\vspace{0.4cm}
\caption{\label{fig:figure3}Input and output wavefunction for a two-photon input pulse with a pulse length of $T=10/\Gamma$. (a) is a contour plot of the input wavefunction \(\Psi_{\text{A}}(x_{1}) \cdot \Psi_{\text{A}}(x_{2})\). (b) is the cross section of the contour plot at $\left|x_{1}-x_{2}\right|=c\tau$, where the time differences are $\tau=0$ (solid line), $\tau=1.4/\Gamma$ (broken line) and $\tau=5/\Gamma$ (dotted line). The horizontal axis \((x_{1}+x_{2})/2\) in (b) represents the average position of the two photons. Likewise, (c) is a contour plot of the output wavefunction \(\Psi_{\text{out}}(x_{1},x_{2})\) and (d) shows cross sections of the contour plot at \(\left|x_{1}-x_{2}\right|=c\tau\), where the time differences are $\tau=0$ (solid line), $\tau=1.4/\Gamma$ (broken line), and $\tau=5/\Gamma$ (dotted line).}
\end{figure}

Figure.~\ref{fig:figure3} (a) shows the contour plot of a long two-photon input pulse \(\Psi_{\text{A}}(x_{1}) \cdot \Psi_{\text{A}}(x_{2})\). The probability amplitude increases from black to white shading. As before, the input pulse has a length of $T=10/\Gamma$. Figure.~\ref{fig:figure3} (b) shows the cross-section of the input wavefunction at $\left|x_{1}-x_{2}\right|=c\tau$ for time differences of $\tau=0,\ \tau=1.4/\Gamma,\ \text{and}\ \tau=5/\Gamma$ between the two photons. Note that the horizontal axis gives the average position of the two photons. The position of the peaks then defines the average delay of both photons at a given distance of $c\tau$ between the two photons.

Figure.~\ref{fig:figure3} (c) and figure.~\ref{fig:figure3} (d) show the output wavefunction \(\Psi_{\text{out}}(x_{1},x_{2})\). A remarkable new feature of this output wavefunction is the negative amplitude region around $x_{1}=x_{2}$. The output amplitude changes from negative to positive values at about $\left|x_{1}-x_{2}\right|=1.4c/\Gamma$. When going further away from $x_{1}=x_{2}$, the spatial features of the output wavefunction eventually become similar to that of the input wavefunction. 

In order to illustrate these spatial features in detail, figure.~\ref{fig:figure3} (d) shows the cross-sections of the output wavefunction at $\left|x_{1}-x_{2}\right|=c\tau$ for $\tau = \tau=0, \tau=1.4/\Gamma,\ \text{and}\ \tau=5/\Gamma$. The negative amplitude of the cross section at a time difference of $\tau=0$ (solid line) is about 3 times as large as the amplitude of the input wavefunction at $\tau=0$ shown in figure.~\ref{fig:figure3} (b) (solid line). Moreover, the peak of the output at $\tau=0$ is delayed by about $2/(3\Gamma)$ compared to the peak of the input wavefunction at the same time difference. The cross-section at a time difference of $\tau=1.4/\Gamma$ (broken line) is nearly zero. This means that the probability density of detecting one photon at a time difference of $\tau=1.4/\Gamma$ after detecting the other photon is vanishingly small at all times. The cross-section at $\tau=5/\Gamma$ (dotted line) is similar to the cross-section of the input wavefunction at the same time difference between the photons. However the output at $\tau=5/\Gamma$ is delayed compared to the input wavefunction by about $2/\Gamma$, which is equal to the delay time observed in the one photon case.

\begin{figure}[htbp]
\psfrag{x1}[t][t]{$x_{1}\ (\text{in units of}\ c/\Gamma)$}
\psfrag{x2}[b][b]{$x_{2}\ (\text{in units of}\ c/\Gamma)$}
\psfrag{x3}[t][t]{$x_{1}$}
\psfrag{x4}[t][t]{$x_{1}$}
\psfrag{20}[tr][tr]{\footnotesize 20}
\psfrag{10}[tr][tr]{\footnotesize 10}
\psfrag{0}[tr][tr]{\footnotesize 0}
\psfrag{-10}[tr][tr]{\footnotesize -10}
\psfrag{-20}[tr][tr]{\footnotesize -20}
\begin{minipage}{.45\linewidth}
\begin{picture}(0,0)
\put(-5,200){(a)}
\end{picture}
\includegraphics[width=\linewidth,height=\linewidth]{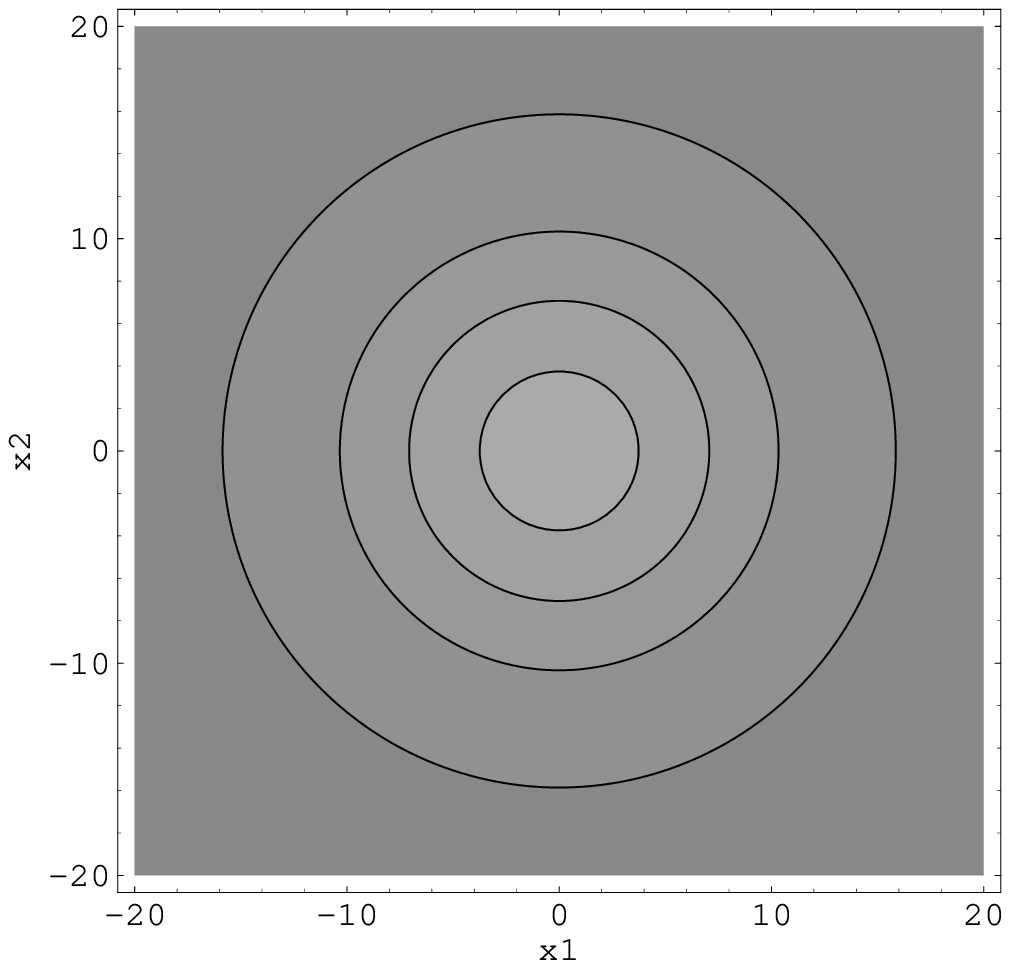}
\end{minipage}
\ \ \ \ \ \ \ \ \ \ \ \ 
\vspace{1cm}
\begin{minipage}{.45\linewidth}
\begin{picture}(0,0)
\put(-5,200){(b)}
\end{picture}
\includegraphics[width=\linewidth,height=\linewidth]{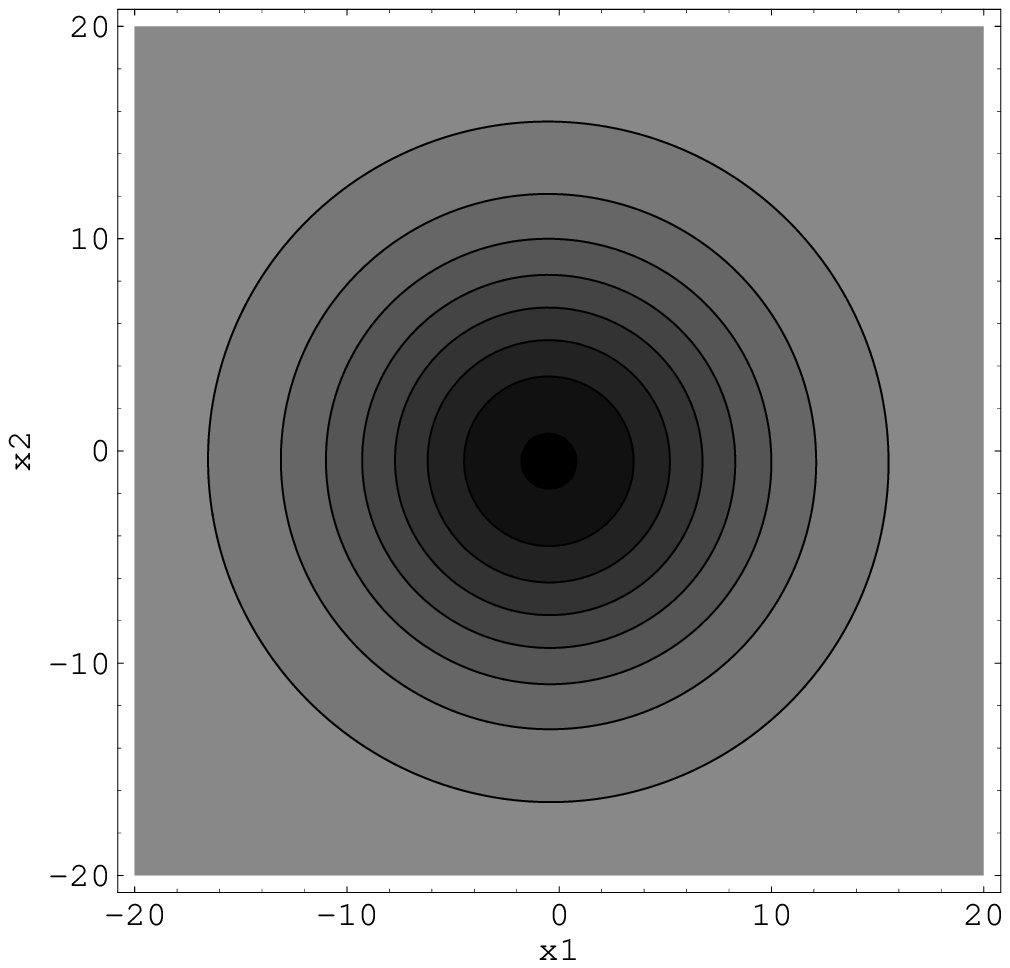}
\end{minipage}
\vspace{1cm}
\begin{minipage}{.45\linewidth}
\begin{picture}(0,0)
\put(-5,200){(c)}
\end{picture}
\includegraphics[width=\linewidth,height=\linewidth]{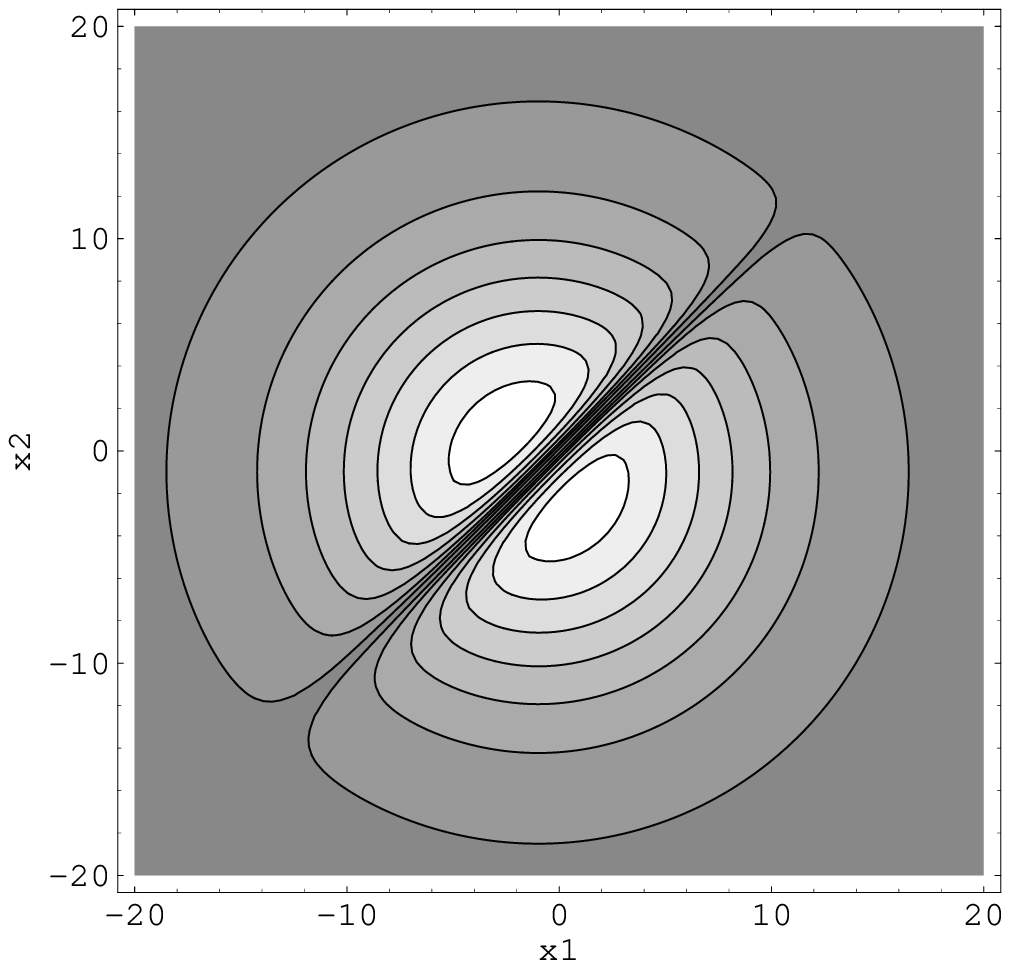}
\end{minipage}\\
\begin{minipage}{.45\linewidth}
\begin{picture}(0,0)
\put(-10,35){(d)}
\put(-10,-10){-0.35}
\put(190,-10){0.3}
\put(110,-10){0}
\end{picture}
\includegraphics[width=\linewidth,height=1cm]{contours.eps}
\end{minipage}
\caption{\label{fig:figure4} Contour plots of the two photon processes $\Psi^{(1)\ \text{to}\ (3)}_{\text{out}}(x_{1},x_{2})$. (a) corresponds to the transmission process, (b) corresponds to the one photon absorption process, (c) corresponds to the two photon absorption process.}
\end{figure}

Let us now consider why such features appear. The two photon case can be understood as a superposition of three processes in the same way that the one photon case can be understood as a superposition of two processes. Figure.\ref{fig:figure4} shows the contour plots of the three components given by eq.(\ref{eq:outputcomponents}). Figure.~\ref{fig:figure4} (a) corresponds to the transmission process, figure.~\ref{fig:figure4} (b) corresponds to the one photon absorption process, and figure.~\ref{fig:figure4} (c) corresponds to the two photon absorption process, respectively. The component of the transmission process (figure.~\ref{fig:figure4} (a)) is identical with the Gaussian input wavefunction, just as in the one photon case, $\Psi^{\text{(1)}}_{\text{out}}(x_{1},x_{2})=\Psi_{\text{A}}(x_{1}) \cdot \Psi_{\text{A}}(x_{2})$. Likewise, the component of the one photon absorption process (figure.~\ref{fig:figure4} (b)), although not a precise Gaussian, can be approximated by the products of single photon transmission and single photon absorption,
\begin{eqnarray}
 \Psi^{\text{(2)}}_{\text{out}}(x_{1},x_{2}) &\approx& \left( -2 \Psi_{\text{A}}(x_{1}+c/\Gamma) \right) \cdot \Psi_{\text{A}}(x_{2}) + \Psi_{\text{A}}(x_{1}) \cdot \left( -2 \Psi_{\text{A}}(x_{2}+c/\Gamma) \right),\nonumber \\
&\approx& -4 \Psi_{\text{A}}(x_{1}) \cdot \Psi_{\text{A}}(x_{2})-2\frac{c}{\Gamma}\left(\frac{\partial}{\partial x_{1}} \Psi_{\text{A}}(x_{1}) \cdot \Psi_{\text{A}}(x_{2}) + \Psi_{\text{A}}(x_{1}) \cdot \frac{\partial}{\partial x_{2}} \Psi_{\text{A}}(x_{2})\right) \nonumber \\
&\approx&  -4 \Psi_{\text{A}}(x_{1}+c/2\Gamma) \cdot \Psi_{\text{A}}(x_{2}+c/2\Gamma) \label{eq:appprocess2}
\end{eqnarray}
This approximation indicates that the negative amplitude of the one photon absorption process in figure.~\ref{fig:figure4} (b) is about 4 times as large as the positive amplitude of the transmission process in figure.~\ref{fig:figure4} (a). Moreover, the wavepacket of the one photon absorption process is delayed by $c/(2\Gamma)$ compared to the transmission process.

The component of the two photon absorption process (figure.~\ref{fig:figure4} (c)) includes the non-linear interaction between the two photons. For $\left|x_{1}-x_{2}\right| \gg c/\Gamma$, it resembles the shifted Gaussian expected for separate one photon absorption processes (see eq.(\ref{eq:appabs2})),
\begin{eqnarray}
\Psi^{(3)}(x_{1},x_{2}) &\approx& 4 \Psi_{\text{A}}(x_{1}+c/\Gamma) \cdot \Psi_{\text{A}}(x_{2}+c/\Gamma).  \label{eq:shiftedgaussian} 
\end{eqnarray}
However, the saturation of the atom completely suppresses this process for $\left|x_{1}-x_{2}\right|=0$. Closer inspection shows that the transition between $\left|x_{1}-x_{2}\right|=0$ and $\left|x_{1}-x_{2}\right| \gg c/\Gamma$ can be approximated by an exponential function of $\left|x_{1}-x_{2}\right|$,
\begin{eqnarray}
\Psi^{\text{(3)}}_{\text{out}}(x_{1},x_{2}) &\approx& 4 \Psi_{\text{A}}(x_{1}+c/\Gamma) \cdot \Psi_{\text{A}}(x_{2}+c/\Gamma) \left(1-e^{-\frac{\Gamma}{c}\left|x_{1}-x_{2}\right|}\right) \label{eq:appprocess3}
\end{eqnarray}
The combination of the two photon absorption process is therefore directly dependent on the time difference $\tau=\left|x_{1}-x_{2}\right|/c$ between the two photons.

\begin{figure}[htbp]
\psfrag{x1}[t][t]{$\left(x_{1}+x_{2}\right)/2\ \ (\text{in units of}\ c/\Gamma)$}
\psfrag{x2}[t][t]{$\left(x_{1}+x_{2}\right)/2\ \ (\text{in units of}\ c/\Gamma)$}
\psfrag{x3}[t][t]{$\left(x_{1}+x_{2}\right)/2\ \ (\text{in units of}\ c/\Gamma)$}
\psfrag{x4}[t][t]{$\left(x_{1}+x_{2}\right)/2\ \ (\text{in units of}\ c/\Gamma)$}
\psfrag{amp1}[b][b]{\resizebox{0.8\hsize}{!}{$\Psi^{(1)\ \text{to}\ (3)}_{\text{out}}(x_{1},x_{2}=x_{1})\ (\text{in units of}\ \Gamma/c)$}}
\psfrag{amp4}[b][b]{\resizebox{0.9\hsize}{!}{$\Psi^{(1)\ \text{to}\ (3)}_{\text{out}}(x_{1},x_{2}=x_{1}+1.4c/\Gamma)\ (\text{in units of}\ \Gamma/c)$}}
\psfrag{amp2}[b][b]{\resizebox{0.9\hsize}{!}{$\Psi^{(1)\ \text{to}\ (3)}_{\text{out}}(x_{1},x_{2}=x_{1}+1.4c/\Gamma)\ (\text{in units of}\ \Gamma/c)$}}
\psfrag{amp3}[b][b]{\resizebox{0.9\hsize}{!}{$\Psi^{(1)\ \text{to}\ (3)}_{\text{out}}(x_{1},x_{2}=x_{1}+5c/\Gamma)\ (\text{in units of}\ \Gamma/c)$}}
\begin{minipage}{.45\linewidth}
\begin{picture}(0,0)
  \put(-5,215){(a)}
\end{picture}
\includegraphics[width=\linewidth,height=\linewidth]{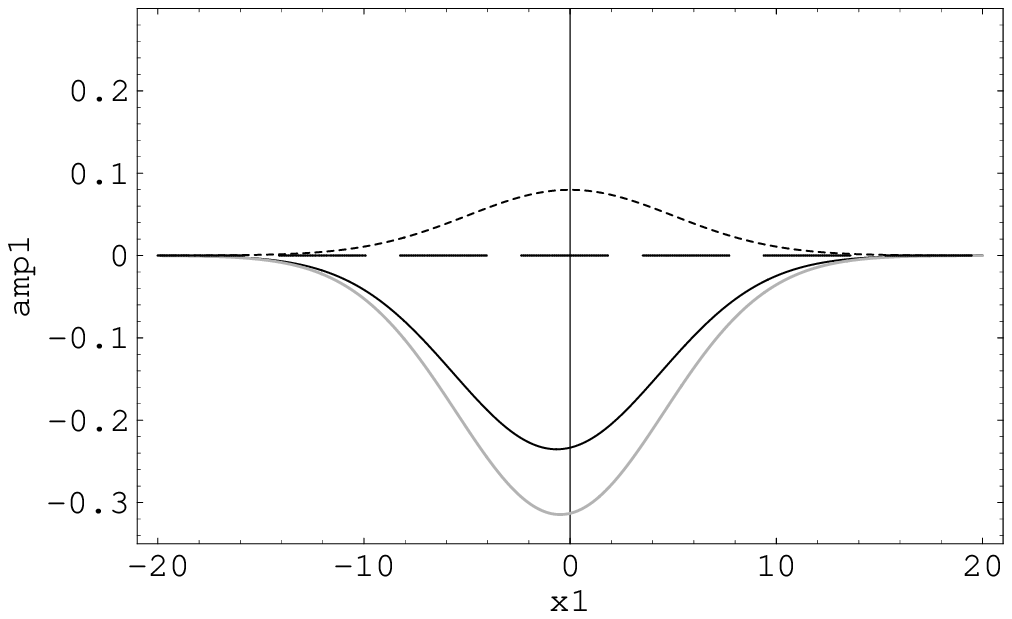}
\end{minipage}
\ \ \ \ \ \ \ \ \ \
\vspace{1cm}
 \begin{minipage}{.45\linewidth}
\begin{picture}(0,0)
  \put(-5,215){(b)}
\end{picture}
\includegraphics[width=\linewidth,height=\linewidth]{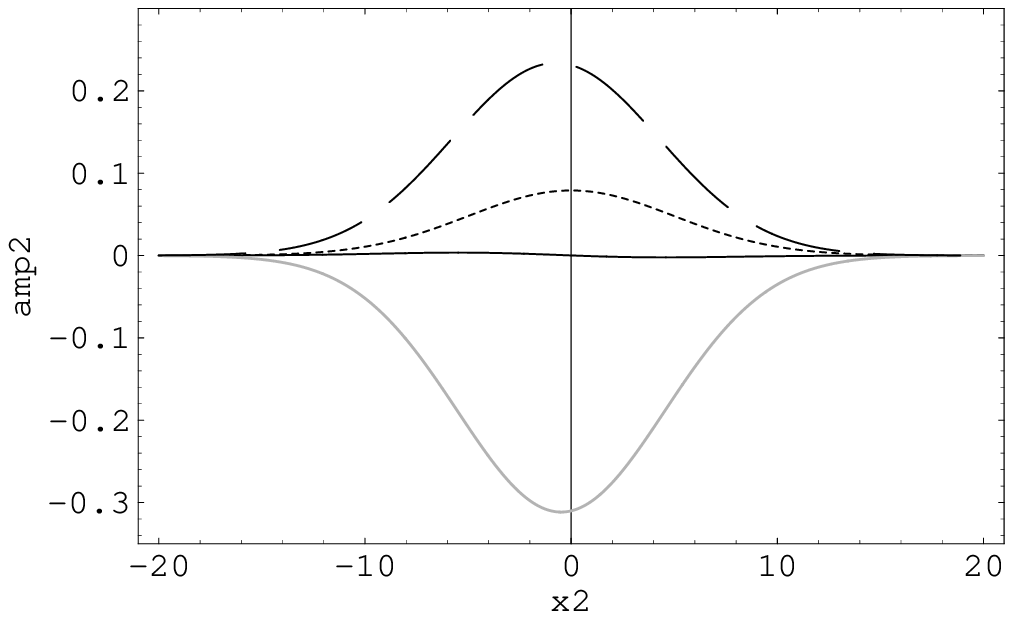}
\end{minipage}
\begin{minipage}{.45\linewidth}
\begin{picture}(0,0)
\put(-5,215){(c)}
\end{picture}
\includegraphics[width=\linewidth,height=\linewidth]{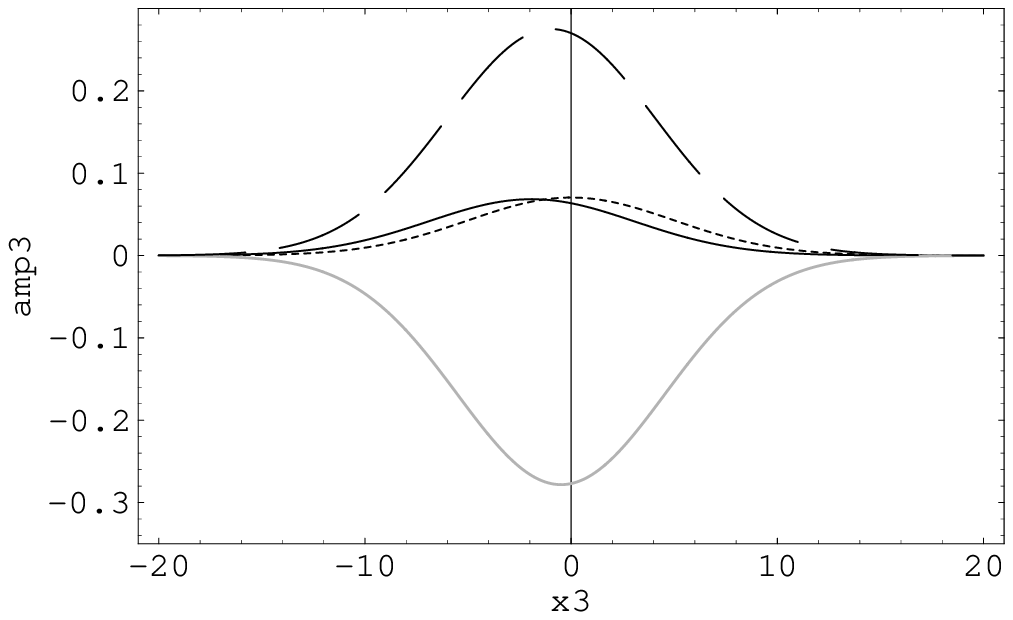}
\end{minipage}
\caption{\label{fig:figure5} Cross-sections of the output amplitude $\Psi_{\text{out}}(x_{1},x_{2})$ and its components $\Psi^{(1)\ \text{to}\ (3)}_{\text{out}}(x_{1},x_{2})$ at time differences $\tau=\left|x_{1}-x_{2}\right|/c$ of (a) $\tau=0$, (b) $\tau=1.4/\Gamma$ and (c) $\tau=5/\Gamma$. The solid lines correspond to the total output amplitude $\Psi_{\text{out}}(x_{1},x_{2})$, the dotted lines correspond to the transmission amplitude $\Psi^{\text{(1)}}_{\text{out}}(x_{1},x_{2})$, the thin lines correspond to the one photon absorption amplitude $\Psi^{\text{(2)}}_{\text{out}}(x_{1},x_{2})$, and the broken lines correspond to the two photon absorption amplitude  $\Psi^{\text{(3)}}_{\text{out}}(x_{1},x_{2})$.}
\end{figure}

In order to illustrate the dependence of the output component on the relative distance between two photons in detail, figure.~\ref{fig:figure5} shows the cross-sections at $\left|x_{1}-x_{2}\right|=c\tau$ of the output amplitude $\Psi_{\text{out}}(x_{1},x_{2})$ (solid lines) and the components of the transmission process $\Psi^{\text{(1)}}_{\text{out}}(x_{1},x_{2})$ (dotted lines), the one photon absorption process $\Psi^{\text{(2)}}_{\text{out}}(x_{1},x_{2})$ (thin lines), and the two photon absorption process $\Psi^{\text{(3)}}_{\text{out}}(x_{1},x_{2})$ (broken lines). The time differences are $\tau=0\ \text{in (a)},\ \tau=1.4/\Gamma\ \text{in (b),}\ \text{and}\ \tau=5/\Gamma\ \text{in (c)}$.
For all three time differences $\tau$, the amplitude of the component of the one photon absorption process (thin lines) is about 4 times as large as the amplitude of the corresponding component of the transmission process (dotted lines). On the other hand, the ratio of the amplitude of the component of the two photon absorption process (broken lines) to the amplitude of the corresponding component of the transmission process (dotted lines) changes depending on the time difference $\tau$ between the two photons as discussed in the following.

Figure.~\ref{fig:figure5} (a) shows the cross sections of the output components at $\tau=0$. In this case, the amplitude of the component of the two photon absorption process, $\Psi_{\text{out}}^{\text{(3)}}(x_{1},x_{2})$, is exactly zero (see eq.(\ref{eq:appprocess3})).
The output amplitude is then given by the sum of the transmission amplitude $\Psi^{\text{(1)}}_{\text{out}}(x_{1},x_{2})=\Psi_{\text{A}}(x_{1}) \cdot \Psi_{\text{A}}(x_{2})$ and the one photon absorption amplitude $\Psi^{\text{(2)}}_{\text{out}}(x_{1},x_{2}) \approx -4 \Psi_{\text{A}}(x_{1}) \cdot \Psi_{\text{A}}(x_{2})$,
\begin{eqnarray}
\Psi_{\text{out}}(x_{1},x_{2}) &\approx& \underbrace{\Psi_{\text{A}}(x_{1}) \cdot \Psi_{\text{A}}(x_{2})}_{\text{transmission}}\ \underbrace{-\ 4 \Psi_{\text{A}}(x_{1}) \cdot \Psi_{\text{A}}(x_{2})}_{\text{one photon absorption}} \nonumber\\
&=& -3 \Psi_{\text{A}}(x_{1}) \cdot \Psi_{\text{A}}(x_{2}). \label{eq:tauzerocase}
\end{eqnarray}
 The amplitude of the output at the time difference of $\tau=0$ (solid line in figure.~\ref{fig:figure5} (a)) is therefore about 3 times as large as the corresponding amplitude of the input.

Figure.~\ref{fig:figure5} (b) shows the cross sections of the output components at $\tau=1.4/\Gamma$. At this special delay time, the factor of $\left(1-e^{-\frac{\Gamma}{c} \left|x_{1}-x_{2}\right|}\right)$ describing the effect of the nonlinearity in eq.(\ref{eq:appprocess3}) is approximately equal to $3/4$. Therefore, the two photon absorption component in the total output can be approximated by $\Psi^{\text{(3)}}(x_{1},x_{2}) \approx 3 \Psi_{\text{A}}(x_{1}) \cdot \Psi_{\text{A}}(x_{2})$. The output wavefunction is then given by
\begin{eqnarray}
\Psi_{\text{out}}(x_{1},x_{2}) &\approx& \underbrace{\Psi_{\text{A}}(x_{1}) \cdot \Psi_{\text{A}}(x_{2})}_{\text{transmission}}\ \underbrace{\ -\ 4 \Psi_{\text{A}}(x_{1}) \cdot \Psi_{\text{A}}(x_{2})}_{\text{one photon absorption}} + \underbrace{3 \Psi_{\text{A}}(x_{1}) \cdot \Psi_{\text{A}}(x_{2})}_{\text{two photon absorption}} \approx 0.
\end{eqnarray}
At a delay time of $\tau=1.4/\Gamma$, the positive amplitudes of the transmission process and the two photon absorption process thus compensate the negative amplitude of the one photon absorption process, resulting in a total output amplitude of nearly zero.

Figure.~\ref{fig:figure5} (c) shows the cross sections of the output components at $\tau=5/\Gamma$. In this case, the amplitude of the component of the two photon absorption process, $\Psi_{\text{out}}^{\text{(3)}}(x_{1},x_{2})$, can be approximated as $4 \Psi_{\text{A}}(x_{1}) \cdot \Psi_{\text{A}}(x_{2})$ since the time difference $\tau$ is sufficiently larger than $1/\Gamma$. The output amplitude is then given by 
\begin{eqnarray}
\Psi_{\text{out}}(x_{1},x_{2}) &\approx& \underbrace{\Psi_{\text{A}}(x_{1}) \cdot \Psi_{\text{A}}(x_{2})}_{\text{transmission}}\ \underbrace{-\ 4 \Psi_{\text{A}}(x_{1}) \cdot \Psi_{\text{A}}(x_{2})}_{\text{one photon absorption}}\ \underbrace{+\ 4 \Psi_{\text{A}}(x_{1}) \cdot \Psi_{\text{A}}(x_{2})}_{\text{two photon absorption}} \nonumber \\
&\approx& \Psi_{\text{A}}(x_{1}) \cdot \Psi_{\text{A}}(x_{2}).
\end{eqnarray}
The amplitude of the output at the time difference of $\tau=5/\Gamma$ (solid line in figure.~\ref{fig:figure5} (c)) is thus almost identical with the amplitude of the input at the same time difference.

To explain the delay of the output wavefunction, we can include the delay times in the arguments of the approximations given by eq.(\ref{eq:appprocess2}) and eq.(\ref{eq:appprocess3}). For $\tau=0$, the output wavefunction can then be approximated as
\begin{eqnarray}
\Psi_{\text{out}}(x_{1},x_{2}) &\approx& \underbrace{\Psi_{\text{A}}(x_{1}) \cdot \Psi_{\text{A}}(x_{2})}_{\text{transmission}}\ \underbrace{-\ 4 \Psi_{\text{A}}(x_{1}+c/2\Gamma) \cdot \Psi_{\text{A}}(x_{2}+c/2\Gamma)}_{\text{one photon absorption}} \nonumber\\
&\approx& -3 \left( \Psi_{\text{A}}(x_{1}) \cdot \Psi_{\text{A}}(x_{2})+\frac{2c}{3\Gamma} \left(\frac{\partial}{\partial x_{1}} \Psi_{\text{A}}(x_{1}) \cdot \Psi_{\text{A}}(x_{2}) + \Psi_{\text{A}}(x_{1}) \cdot  \frac{\partial}{\partial x_{2}} \Psi_{\text{A}}(x_{2}) \right)\right) \nonumber \\
&\approx& -3 \Psi_{\text{A}}(x_{1}+2c/(3\Gamma)) \cdot \Psi_{\text{A}}(x_{2}+2c/(3\Gamma)). 
\end{eqnarray}
The amplitude of the output wavefunction at a time difference of $\tau=0$ is thus delayed by about $2/(3\Gamma)$ compared to the amplitude of the input wavefunction at the same time difference. In the case of $\tau=1.4/\Gamma$, the output wavefunction is nearly zero, and no delay time can be defined. For $\tau=5/\Gamma$, the two photon component $\Psi^{\text{(3)}}_{\text{out}}(x_{1},x_{2})$ of the output wavefunction can be approximated using the limit of $\left|x_{1}-x_{2}\right| \gg c/\Gamma$ given in eq.(\ref{eq:shiftedgaussian}). Since in this limit, the two photons are absorbed and reemitted independently of each other, the output wavefunction can then be factorized into one photon processes (see eq.(\ref{eq:apponeoutput})),
\begin{eqnarray}
\Psi_{\text{out}}(x_{1},x_{2}) &\approx& \left(\Psi_{\text{A}}(x_{1}) - 2 \Psi_{\text{A}}(x_{1}+c/\Gamma) \right) \cdot \left( \Psi_{\text{A}}(x_{2}) - 2 \Psi_{\text{A}}(x_{2}+c/\Gamma)\right) \nonumber \\
&\approx& \left(-\Psi_{\text{A}}(x_{1}+2c/\Gamma)\right) \cdot \left(-\Psi_{\text{A}}(x_{2}+2c/\Gamma) \right).
\end{eqnarray}
The delay of the peak of the output wavefunction for $\left|x_{1}-x_{2}\right| \gg c/\Gamma$ is thus equal to the delay for the one photon output wavefunction discussed in section \ref{sec:level3}.1. In summary, the delay of the negative amplitude at $x_{1}=x_{2}$ is equal to $1/3$ of the delay of the positive amplitude at $\left|x_{1}-x_{2}\right| \gg c/\Gamma$. This means that the effect of the non-linear interaction between the photons at $x_{1}=x_{2}$ reduces the total delay of the output wavefunction as compared to the delay caused by the uncorrelated transmission and absorption processes at $\left|x_{1}-x_{2}\right| \gg c/\Gamma$.
\section{RESPONSE TO A SHORT PULSE \label{sec:level4}}
As shown in the long pulse case, the atomic nonlinearity is characterized by a negative amplitude of the output wavefunction around $x_{1}=x_{2}$. In the long pulse case, this nonlinear effect is limited to a region of $\left|x_{1}-x_{2}\right|<1.4c/\Gamma$. In order to realize a non-linear phase change throughout the pulse, it is necessary to shorten the pulse to a pulse length comparable to $1/\Gamma$. In the following, we therefore discuss the response of the atom to a pulse of length $T=1/\Gamma$. 
\subsection{Explanation of the output one-photon wavepacket in terms of one photon processes}
\begin{figure}[htbp]
\psfrag{x}[t][t]{$x\ (\text{in units of}\ c/\Gamma)$}
\psfrag{amp}[b][b]{$\Psi_{\text{in/out}}(x)\ (\text{in units of }\Gamma/c)$}
\psfrag{amp1}[b][b]{$\Psi^{\text{(I)},\ \text{(II)}}_{\text{out}}(x)\ (\text{in units of}\ \Gamma/c)$}
\begin{minipage}{.45\linewidth}
\begin{picture}(0,0)
\put(-5,200){(a)}
\end{picture}
\includegraphics[width=\linewidth,height=\linewidth]{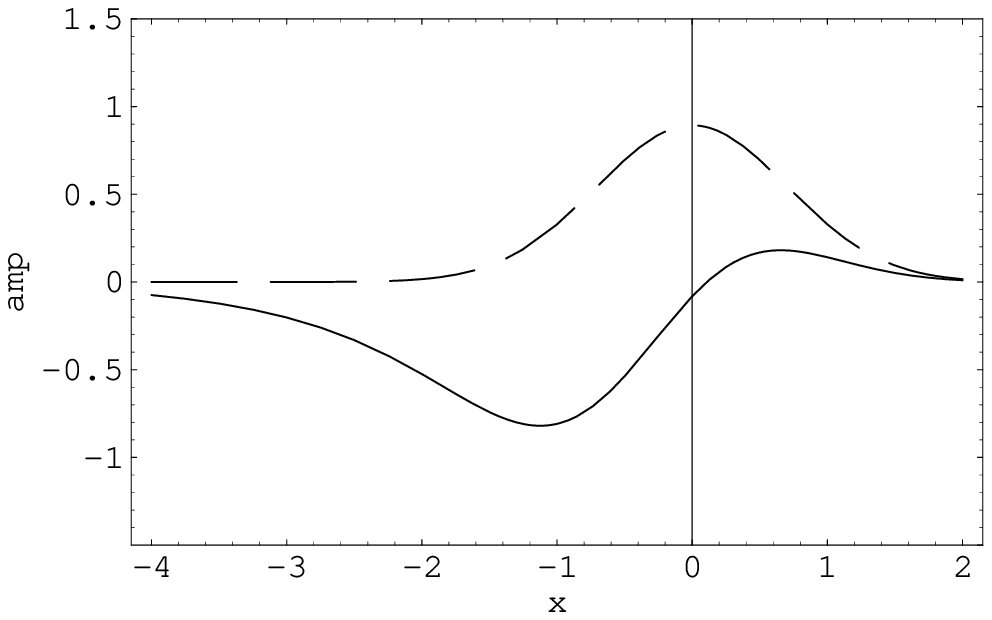}
\end{minipage}
\ \ \ \ \ \ \ \ \ \ \ \ 
\begin{minipage}{.45\linewidth}
\begin{picture}(0,0)
\put(-5,200){(b)}
\end{picture}
\includegraphics[width=\linewidth,height=\linewidth]{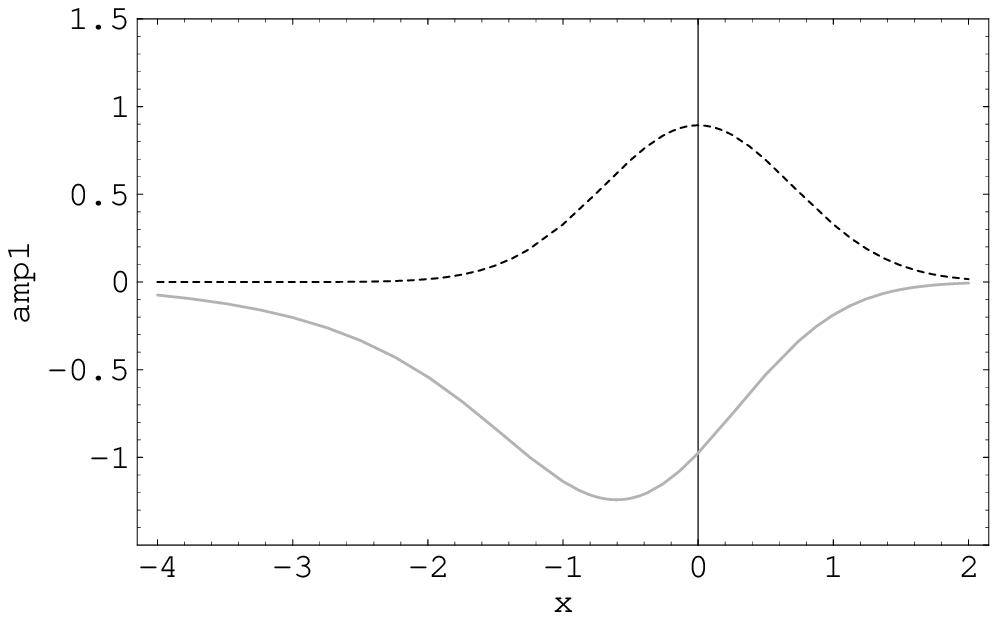}
\end{minipage}
\caption{\label{fig:figure6}Gaussian input and output one-photon wavepackets for an input wavepacket length of \(T=1/\Gamma\). In (a), the broken line corresponds to the input wavefunction $\Psi_{\text{A}}(x)$ and the solid line corresponds to the output wavefunction $\Psi_{out}(x)$. In (b), the dotted line corresponds to the component of the process (I): single photon transmission without absorption and the thin line corresponds to the component of the process (II): single photon reemission after absorption.}
\end{figure}

The line plot in figure.~\ref{fig:figure6} (a) shows the short one-photon pulse (broken line) and the corresponding output wavefunction (solid line). The length of the input wavepacket is \(T=1/\Gamma\).
The pulse shape of the output pulse is now quite different from the input pulse. Specifically, the output amplitude is still positive for $x>0$ and only changes to negative values in the latter part of the pulse. Let us now analyze the origin of these features.

Figure.~\ref{fig:figure6} (b) shows the components corresponding to  different light-atom interaction processes as given by eqs.(\ref{eq:oneoutputcomponents}). The dotted line is the component of single photon transmission without absorption, $\Psi_{\text{prop}}(x)$. The thin line is that of single photon reemission after absorption, $\Psi_{\text{abs}}(x)$. As before, $\Psi_{\text{prop}}(x)=\Psi_{\text{A}}(x)$ according to eqs.(\ref{eq:oneoutputcomponents}).
On the other hand, the absorption component $\Psi_{\text{abs}}(x)$ (thin line) is given by the convolution with the exponential function of the dipole relaxation, given by eq.(\ref{eq:absgenerater}) and eqs.(\ref{eq:oneoutputcomponents}). The peak of this component is shifted by about $0.6/\Gamma$ compared to the component of the transmission process. Furthermore, the peak of the amplitude is only about 1.3 times as large as the corresponding amplitude of the component of the transmission process. Therefore, the front part of the output wavefunction is dominated by the positive component of the transmission process $\Psi_{\text{prop}}(x)$, even though the major part of the output wavefunction is still dominated by the negative amplitude component of the absorption process $\Psi_{\text{abs}}(x)$. The large delay between the components and the reduced amplitude of the dominant negative absorption component thus cause the asymmetry of the output wavefunction seen in figure.~\ref{fig:figure6} (a).

\subsection{Explanation of the two-photon output wavepacket in terms of two photon processes}
\begin{figure}[htbp]
\psfrag{x1}[t][t]{$x_{1}\ (\text{in units of}\ c/\Gamma)$}
\psfrag{x2}[b][b]{$x_{2}\ (\text{in units of}\ c/\Gamma)$}
\psfrag{crossx}[t][t]{$\left(x_{1}+x_{2}\right)/2\ \ (\text{in units of}\ c/\Gamma)$}
\psfrag{crossamp1}[b][b]{$\Psi_{\text{in}}(x_{1},x_{2}=x_{1}+c\tau)\ (\text{in units of}\ \Gamma/c)$}
\begin{minipage}{.45\linewidth}
\psfrag{2}[tr][tr]{\footnotesize 2}
\psfrag{1}[tr][tr]{\footnotesize 1}
\psfrag{0}[tr][tr]{\footnotesize 0}
\psfrag{-1}[tr][tr]{\footnotesize -1}
\psfrag{-2}[tr][tr]{\footnotesize -2}
\psfrag{-3}[tr][tr]{\footnotesize -3}
\psfrag{-4}[tr][tr]{\footnotesize -4}
\begin{picture}(0,0)
\put(-5,200){(a)}
\end{picture}
\includegraphics[width=\linewidth,height=\linewidth]{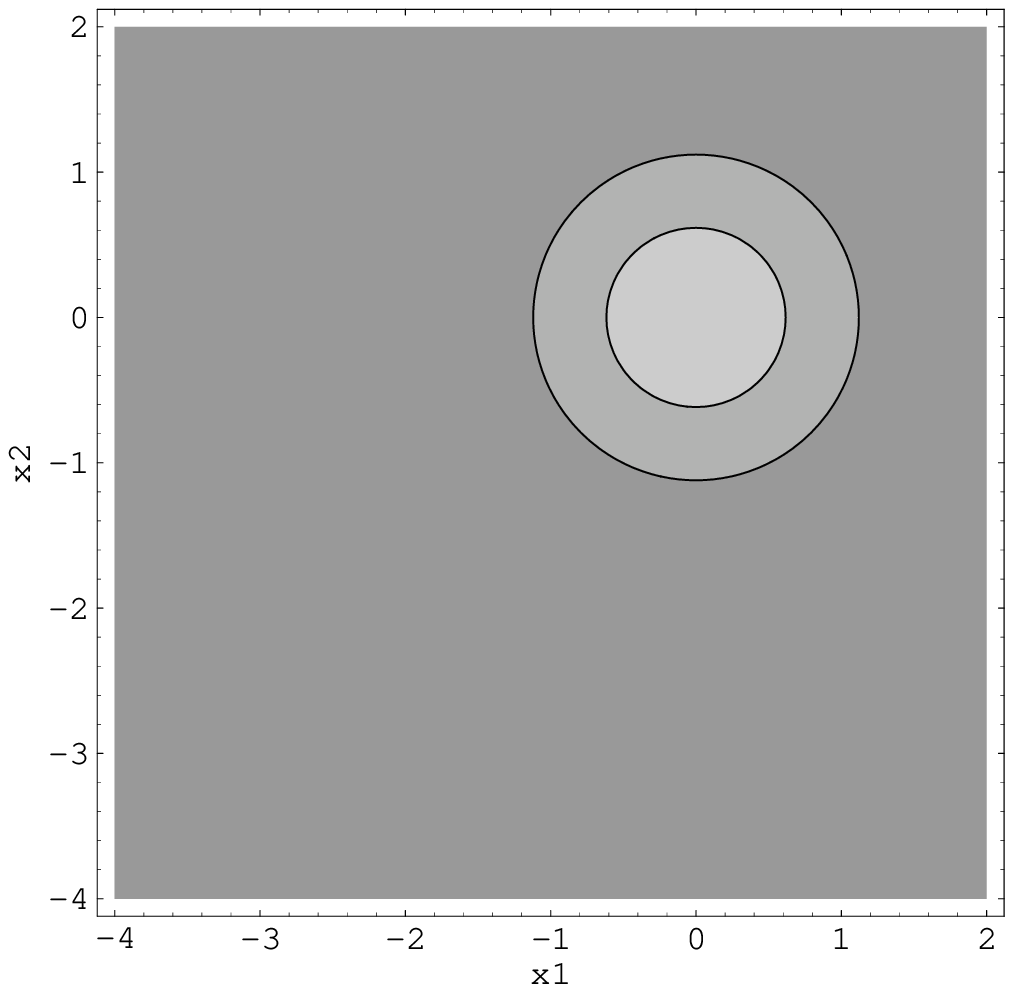}
\end{minipage}
\ \ \ \ \ \ \ \ \ \ \ \
\vspace{1cm} 
\begin{minipage}{.45\linewidth}
\begin{picture}(0,0)
\put(-20,200){(b)}
\end{picture}
\includegraphics[width=\linewidth,height=\linewidth]{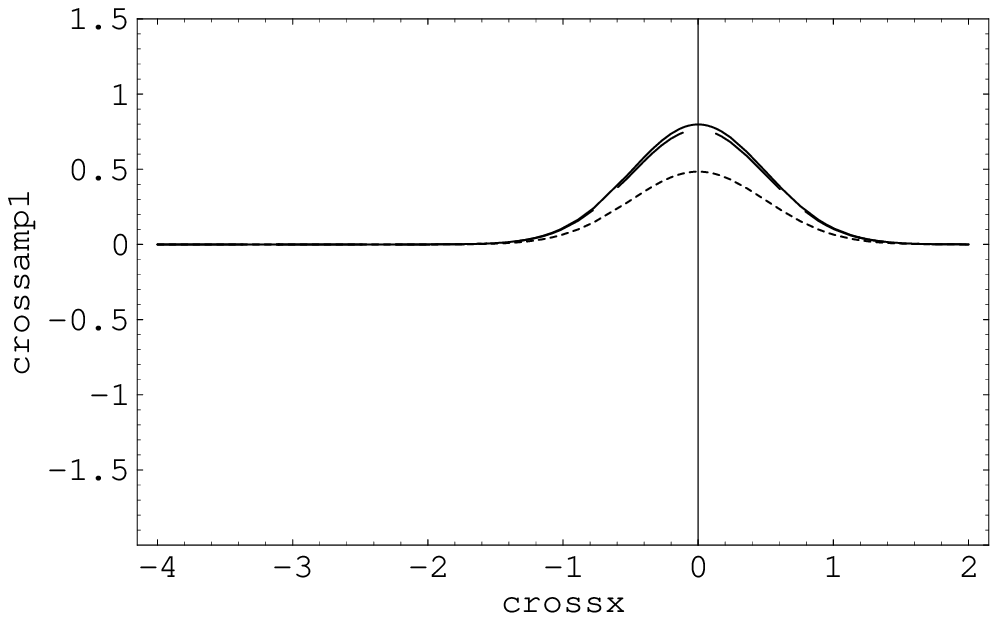}
\end{minipage}
\psfrag{space}[t][t]{$x_{1}\ (\text{in units of}\ c/\Gamma)$}
\psfrag{crossamp2}[b][b]{$\Psi_{\text{out}}(x_{1},x_{2}=x_{1}+c\tau)\ (\text{in units of}\ \Gamma/c)$}
\begin{minipage}{.45\linewidth}
\psfrag{2}[tr][tr]{\footnotesize 2}
\psfrag{1}[tr][tr]{\footnotesize 1}
\psfrag{0}[tr][tr]{\footnotesize 0}
\psfrag{-1}[tr][tr]{\footnotesize -1}
\psfrag{-2}[tr][tr]{\footnotesize -2}
\psfrag{-3}[tr][tr]{\footnotesize -3}
\psfrag{-4}[tr][tr]{\footnotesize -4}
\begin{picture}(0,0)
\put(-5,200){(c)}
\end{picture}
\includegraphics[width=\linewidth,height=\linewidth]{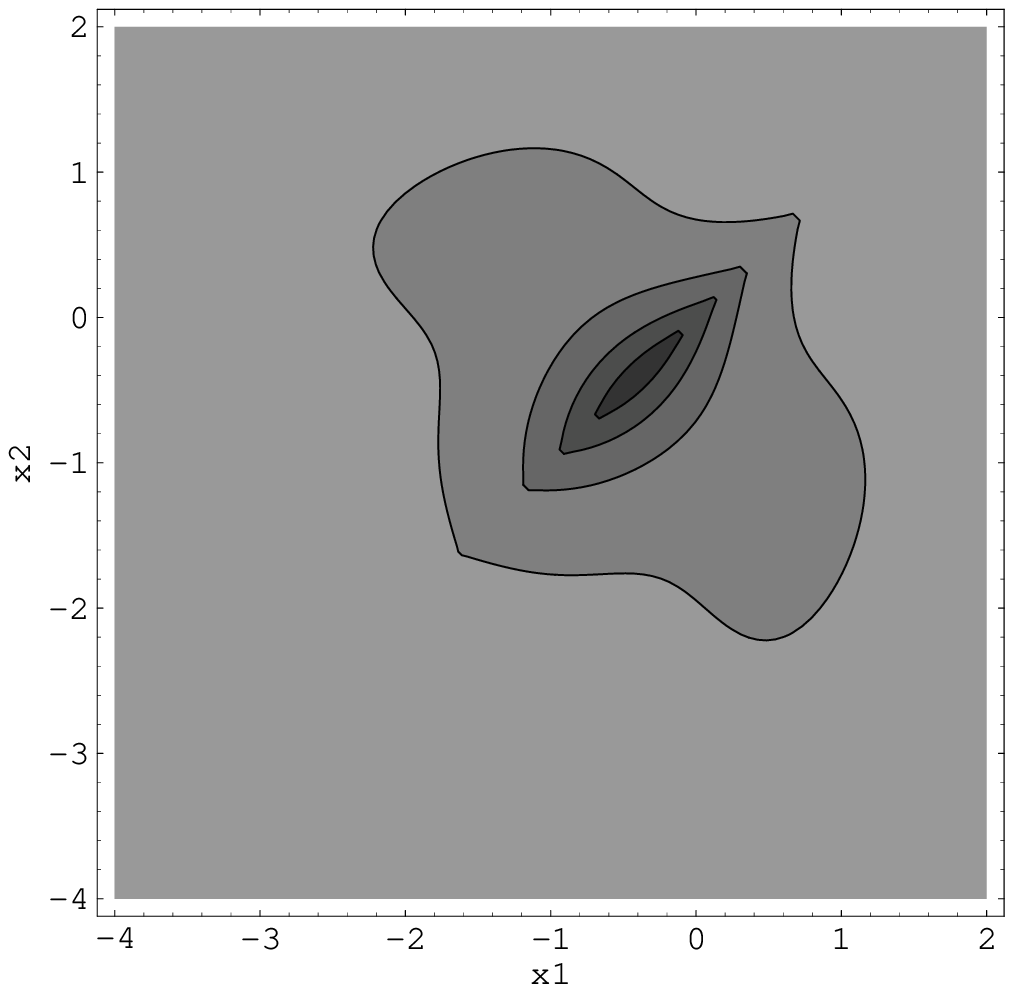}
\end{minipage}
\ \ \ \ \ \ \ \ \ \ \ \
\vspace{1cm} 
\begin{minipage}{.45\linewidth}
\begin{picture}(0,0)
\put(-20,200){(d)}
\end{picture}
\includegraphics[width=\linewidth,height=\linewidth]{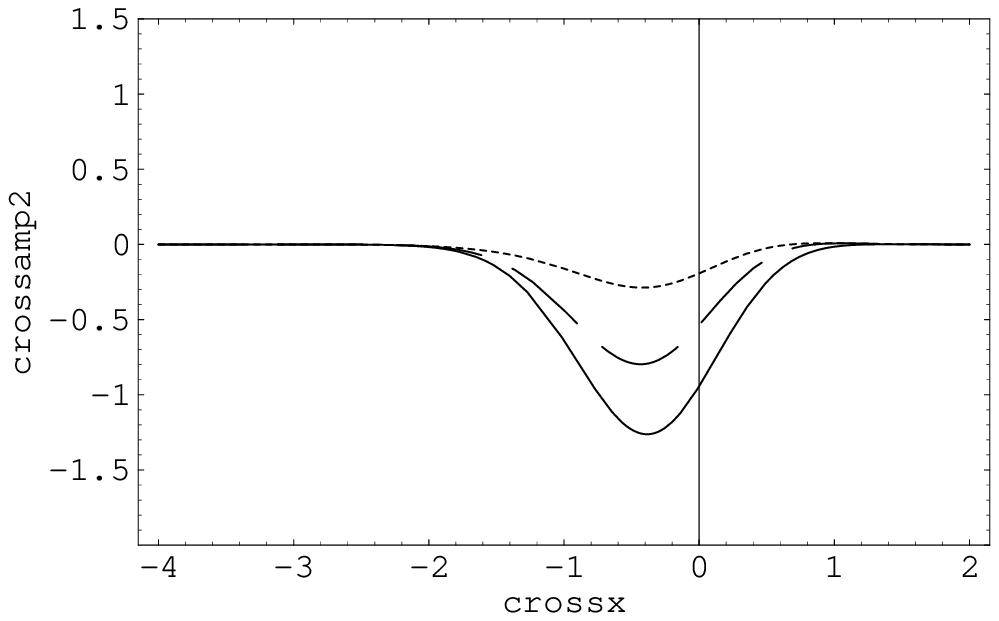}
\end{minipage}
\begin{minipage}{.45\linewidth}
\begin{picture}(0,0)
\put(-15,25){(e)}
\put(-5,-10){-2.0}
\put(185,-10){1.5}
\put(120,-10){0}
\end{picture}
\includegraphics[width=\linewidth,height=1cm]{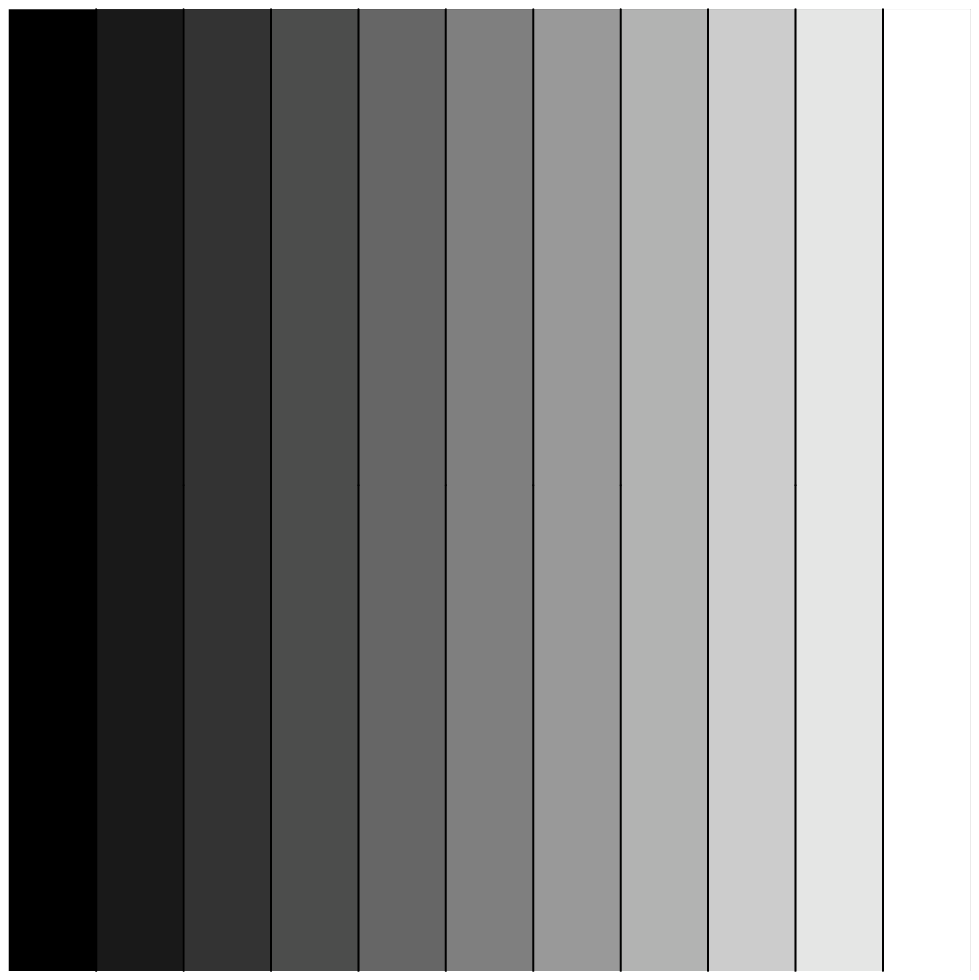}
\end{minipage}
\vspace{0.4cm}
\caption{\label{fig:figure7} Input and output wavefunction for a two-photon input pulse with a pulse length of $T=1/\Gamma$. (a) is a contour plot of the input wavefunction $\Psi_{\text{A}}(x_{1}) \cdot \Psi_{\text{A}}(x_{2})$. (b) is the cross-section of the contour line plot at \(\left|x_{1}-x_{2}\right|=c\tau\), where the time differences are $\tau=0$ (solid line), $\tau=0.3/\Gamma$ (broken line) and $\tau=1/\Gamma$ (dotted line). The horizontal axis $(x_{1}+x_{2})/2$ in (b) is the average position of the two photons. Likewise, (c) is a contour plot of the output wavefunction \(\Psi_{\text{out}}(x_{1},x_{2})\) and (d) shows cross sections of the contour plot at \(\left|x_{1}-x_{2}\right|=c\tau\), where the time differences are $\tau=0$ (solid line), $\tau=0.3/\Gamma$ (broken line) and $\tau=1/\Gamma$ (dotted line).}
\end{figure}

Let us now consider the output wavefunction for the short two photon pulse. Figure.~\ref{fig:figure7} (a) shows the contour plot of the short two-photon input pulse \(\Psi_{\text{A}}(x_{1}) \cdot \Psi_{\text{A}}(x_{2})\). The probability amplitude increases from black to white shading. As before, the input pulse has a length of \(T=1/\Gamma\).
Figure.~\ref{fig:figure7} (b) shows the cross-section of the input wavefunction at $\left|x_{1}-x_{2}\right|=c\tau$ for time differences of $\tau=0$, $\tau=0.3/\Gamma$, and $\tau=1/\Gamma$ between the two photons. Figure.~\ref{fig:figure7} (c) and figure.~\ref{fig:figure7} (d) show the output wavefunction $\Psi_{\text{out}}(x_{1},x_{2})$. The output amplitude is now negative everywhere. However, the pulse shape of the output shown in figure.~\ref{fig:figure7} (c) is quite different from the shape of the input pulse shown in figure.~\ref{fig:figure7} (a). In order to illustrate the spatial features of the output wavefunction in detail, figure.~\ref{fig:figure7} (d) shows the cross-sections of the output wavefunction for time differences of $\tau=0$, $\tau=0.3/\Gamma$, and $\tau=1/\Gamma$ between the two photons. Note that the shape of the cross-sections of the output in figure.~\ref{fig:figure7} (d) is similar to that of the input cross-sections in figure.~\ref{fig:figure7} (b). However, the amplitudes are quite different. Specifically, the output amplitude at a time difference of $\tau=0$ (figure.~\ref{fig:figure7} (d), solid line) is larger than the input amplitude at the same time difference (figure.~\ref{fig:figure7} (b), solid line), while the output amplitude at a time difference of $\tau=1/\Gamma$ (figure.~\ref{fig:figure7} (d), dotted line) is smaller than the input amplitude at the same time difference (figure.~\ref{fig:figure7} (b), dotted line). This indicates that the main difference in pulse shape between the input and output pulses is that the output amplitude decreases faster than the input amplitude between $\tau=0$ and $\tau=c/\Gamma$. On the other hand, the delay time of the peak of the output amplitude does not change much. Its value is about $0.4/\Gamma$ for all time differences $\tau$ between the photons.

\begin{figure}[htbp]
\psfrag{x1}[t][t]{$x_{1}\ (\text{in units of}\ c/\Gamma)$}
\psfrag{x2}[b][b]{$x_{2}\ (\text{in units of}\ c/\Gamma)$}
\psfrag{2}[tr][tr]{\footnotesize 2}
\psfrag{1}[tr][tr]{\footnotesize 1}
\psfrag{0}[tr][tr]{\footnotesize 0}
\psfrag{-1}[tr][tr]{\footnotesize -1}
\psfrag{-2}[tr][tr]{\footnotesize -2}
\psfrag{-3}[tr][tr]{\footnotesize -3}
\psfrag{-4}[tr][tr]{\footnotesize -4}
\begin{minipage}{.45\linewidth}
\begin{picture}(0,0)
\put(-5,200){(a)}
\end{picture}
\includegraphics[width=\linewidth,height=\linewidth]{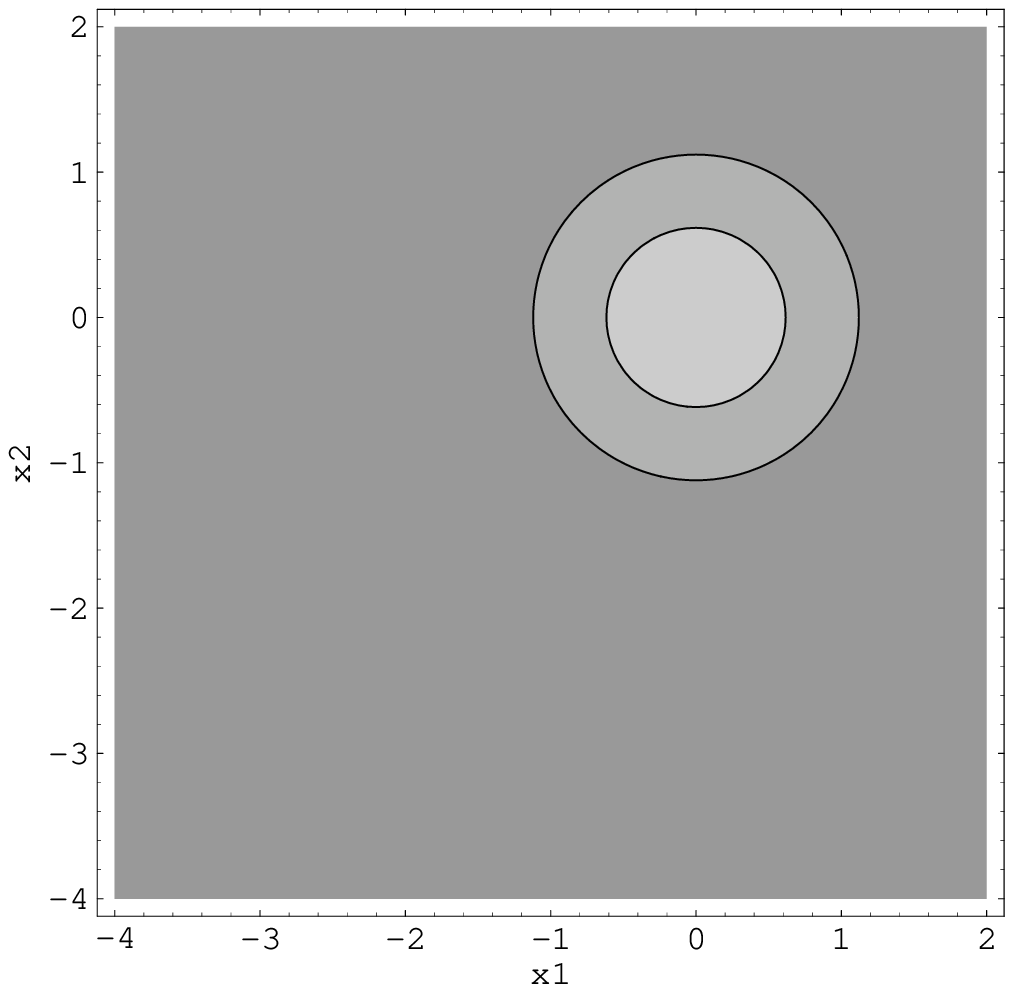}
\end{minipage}
\ \ \ \ \ \ \ \ \ \ \ \
\vspace{1cm} 
\begin{minipage}{.45\linewidth}
\begin{picture}(0,0)
\put(-5,200){(b)}
\end{picture}
\includegraphics[width=\linewidth,height=\linewidth]{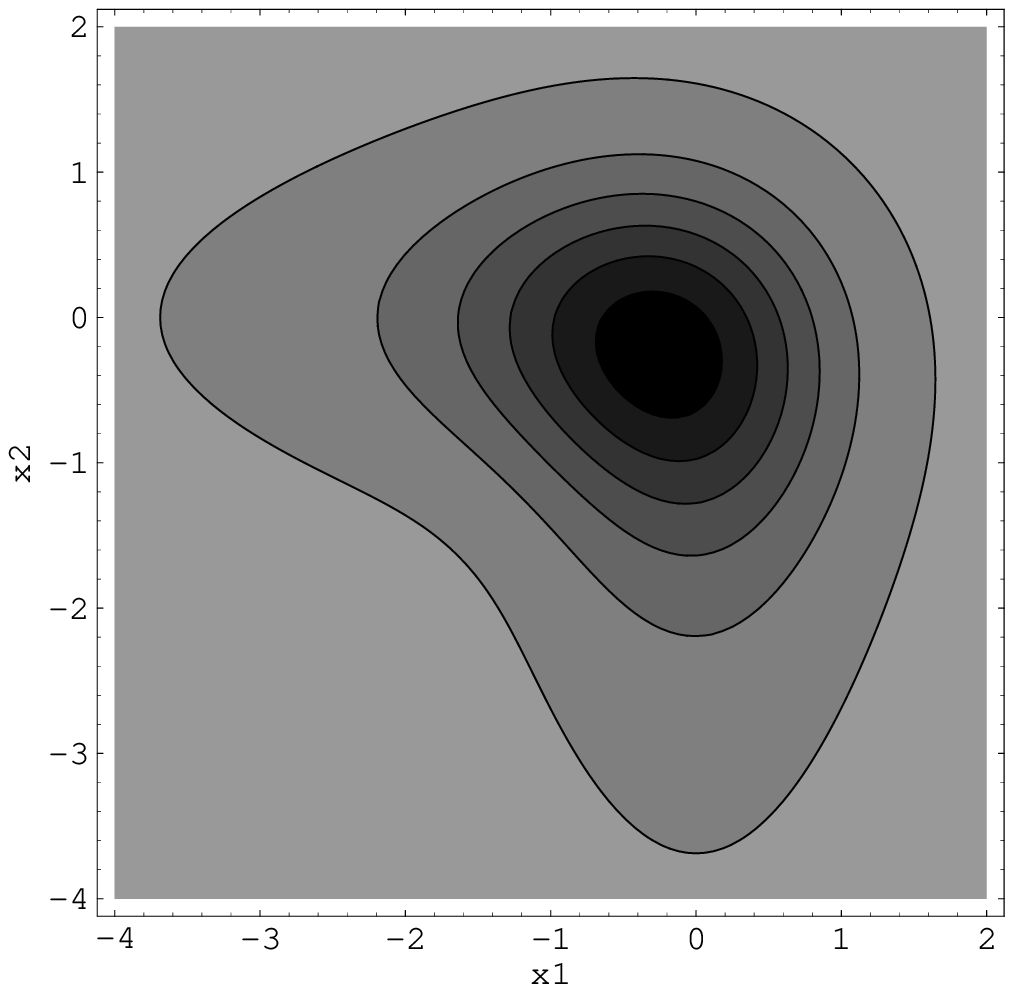}
\end{minipage}
\vspace{1cm}
\begin{minipage}{.45\linewidth}
\begin{picture}(0,0)
\put(-5,200){(c)}
\end{picture}
\includegraphics[width=\linewidth,height=\linewidth]{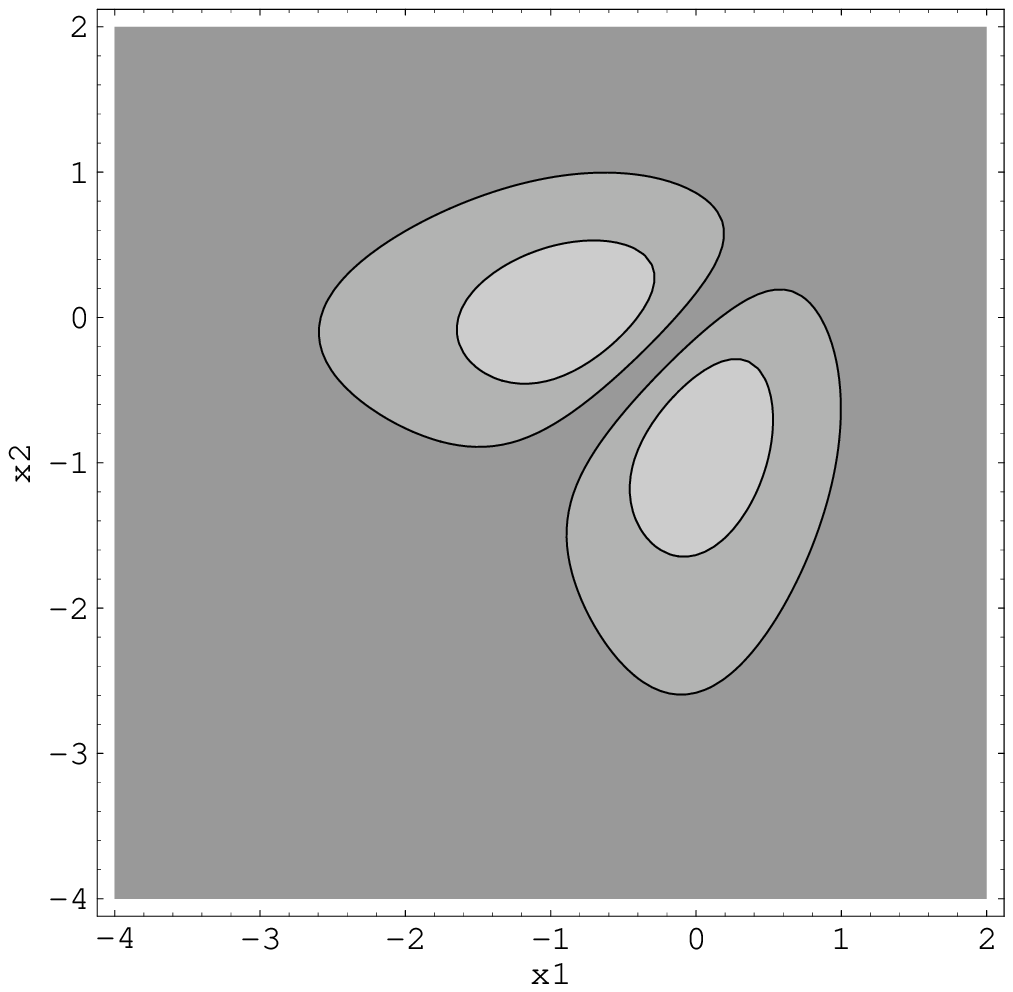}
\end{minipage}\\
\begin{minipage}{.45\linewidth}
\begin{picture}(0,0)
\put(-10,35){(d)}
\put(-10,-10){-0.2}
\put(190,-10){1.5}
\put(120,-10){0}
\end{picture}
\includegraphics[width=\linewidth,height=1cm]{midcontours.eps}
\end{minipage}
\caption{\label{fig:figure8} Contour plots of the two photon processes $\Psi^{\text{(1)}\ \text{to}\ \text{(3)}}_{\text{out}}(x_{1},x_{2})$. (a) corresponds to the transmission process, (b) corresponds to the one photon absorption process, (c) corresponds to the two photon absorption process.}
\end{figure}

Let us now consider why such features appear. Figure.~\ref{fig:figure8} shows the contour plot of the three components given by eq.(\ref{eq:outputcomponents}). Figure.~\ref{fig:figure8} (a) corresponds to the transmission process, figure.~\ref{fig:figure8} (b) corresponds to the one photon absorption process and figure.~\ref{fig:figure8} (c) corresponds to the two photon absorption process. As before, the component of the transmission process (figure.~\ref{fig:figure8} (a)) is identical with the input wavefunction, $\Psi^{\text{(1)}}_{\text{out}}(x_{1},x_{2})=\Psi_{\text{prop}}(x_{1}) \cdot \Psi_{\text{prop}}(x_{2})$. Likewise, the component of the one photon absorption process (figure.~\ref{fig:figure8} (b)) can be described as
\begin{eqnarray}
\Psi^{\text{(2)}}_{\text{out}}(x_{1},x_{2}) &=& \Psi_{\text{prop}}(x_{1}) \cdot \Psi_{\text{abs}}(x_{2}) + \Psi_{\text{abs}}(x_{1}) \cdot \Psi_{\text{prop}}(x_{2})  
\end{eqnarray}
Since the single photon transmission $\Psi_{\text{prop}}(x)$ is positive and the single photon absorption $\Psi_{\text{abs}}(x)$ is negative, the total amplitude of this component is negative for all $x_{1},\ x_{2}$.

The component of the two photon absorption process (figure.~\ref{fig:figure8} (c)) includes the non-linear interaction between the two photons. The positive amplitude of this component is therefore always smaller than the product of the single photon absorption amplitudes, $\Psi_{\text{abs}}(x_{1}) \cdot \Psi_{\text{abs}}(x_{2})$. Specifically, the two photon absorption component $\Psi^{\text{(3)}}_{\text{out}}(x_{1},x_{2})$ is exactly zero at $x_{1}=x_{2}$ and increases rapidly up to a maximal amplitude at about $\left|x_{1}-x_{2}\right|=c/\Gamma$. For larger photon distances, $\left|x_{1}-x_{2}\right|>c/\Gamma$, $\Psi^{\text{(3)}}_{\text{out}}(x_{1},x_{2})$ decreases as a consequence of the decrease in the amplitudes of the photon absorption processes, $\Psi_{\text{abs}}(x_{1}) \cdot \Psi_{\text{abs}}(x_{2})$. Note however that, at its maximal amplitudes near $\left|x_{1}-x_{2}\right|=c/\Gamma$, the two photon absorption amplitude $\Psi^{\text{(3)}}_{\text{out}}(x_{1},x_{2})$ is still significantly lower than the product of the single photon absorption amplitudes, $\Psi_{\text{abs}}(x_{1}) \cdot \Psi_{\text{abs}}(x_{2})$. This means that the non-linear interaction of the photons significantly affects the whole output pulse in the case of an input pulse length of $T=1/\Gamma$.

\begin{figure}[htbp]
\psfrag{x1}[t][t]{$\left(x_{1}+x_{2}\right)/2\ \ (\text{in units of}\ c/\Gamma)$}
\psfrag{amp1}[][]{\resizebox{0.8\hsize}{!}{$\Psi^{(1)\ \text{to}\ (3)}_{\text{out}}(x_{1},x_{2}=x_{1})\ (\text{in units of}\ \Gamma/c)$}}
\psfrag{amp2}[][]{\resizebox{0.9\hsize}{!}{$\Psi^{(1)\ \text{to}\ (3)}_{\text{out}}(x_{1},x_{2}=x_{1}+0.3c/\Gamma)\ (\text{in units of}\ \Gamma/c)$}}
\psfrag{amp3}[][]{\resizebox{0.9\hsize}{!}{$\Psi^{(1)\ \text{to}\ (3)}_{\text{out}}(x_{1},x_{2}=x_{1}+c/\Gamma)\ (\text{in units of}\ \Gamma/c)$}}
\begin{minipage}{.45\linewidth}
\begin{picture}(0,0)
  \put(-5,210){(a)}
\end{picture}
\includegraphics[width=\linewidth,height=\linewidth]{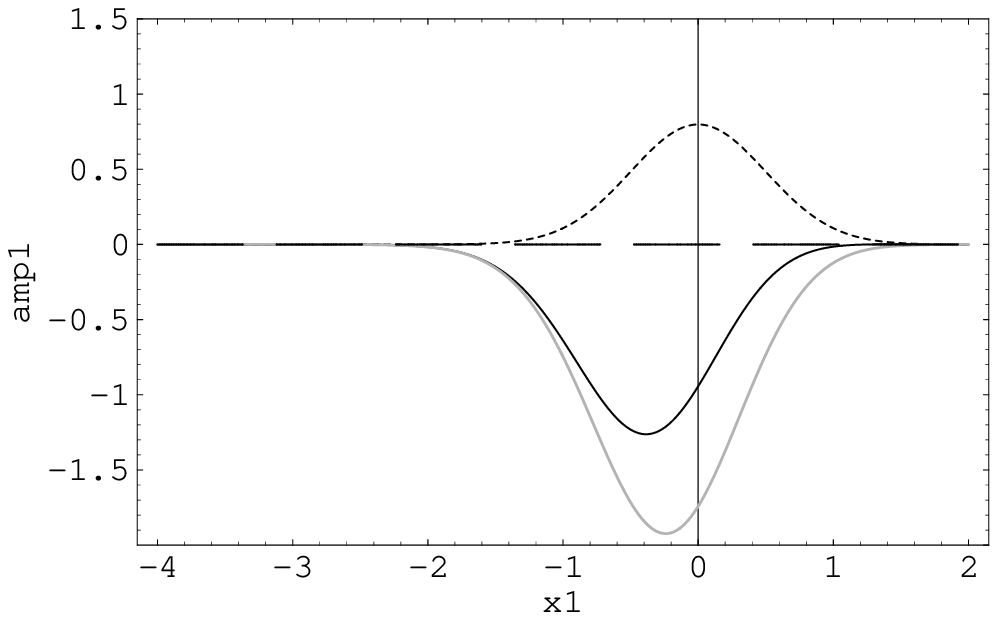}
\end{minipage}
\ \ \ \ \ \ \ \ \ \
\vspace{1cm}
 \begin{minipage}{.45\linewidth}
\begin{picture}(0,0)
  \put(-5,210){(b)}
\end{picture}
\includegraphics[width=\linewidth,height=\linewidth]{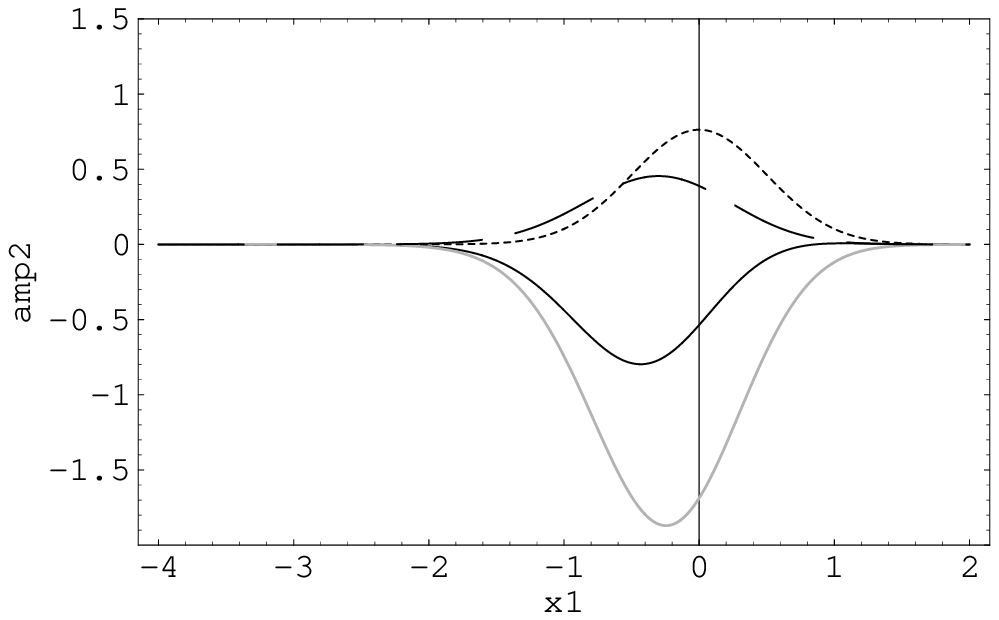}
\end{minipage}\\
\begin{minipage}{.45\linewidth}
\begin{picture}(0,0)
\put(-5,210){(c)}
\end{picture}
\includegraphics[width=\linewidth,height=\linewidth]{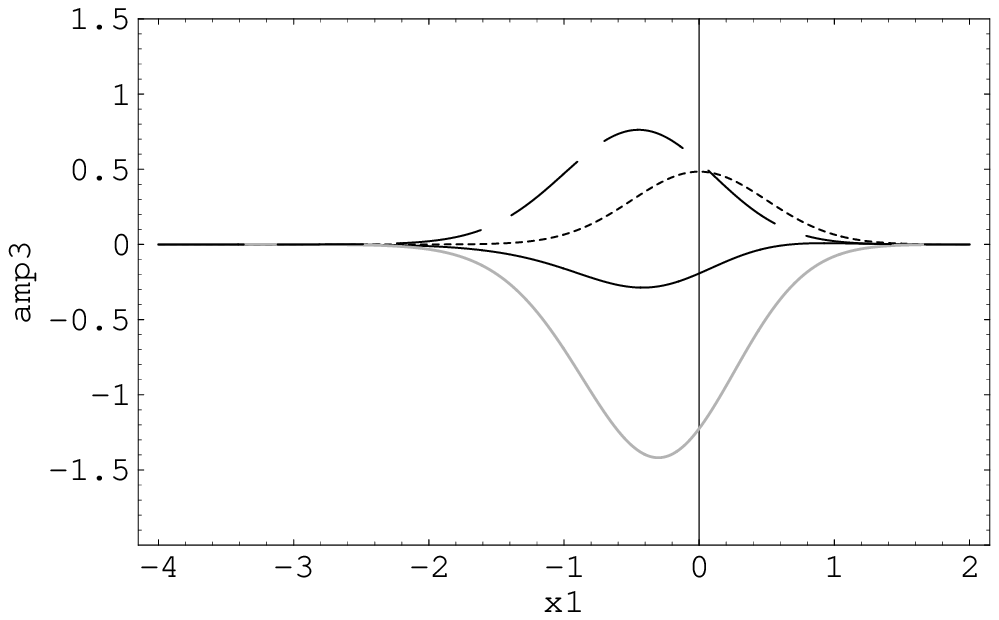}
\end{minipage}
\caption{\label{fig:figure9}
 Cross-sections of the output amplitude $\Psi_{\text{out}}(x_{1},x_{2})$ and its components $\Psi^{(1)\ \text{to}\ (3)}_{\text{out}}(x_{1},x_{2})$ at time differences $\tau=\left|x_{1}-x_{2}\right|/c$ of (a) $\tau=0$, (b) $\tau=0.3/\Gamma$ and (c) $\tau=1/\Gamma$. The solid lines correspond to the total output amplitude $\Psi_{\text{out}}(x_{1},x_{2})$, the dotted lines correspond to the transmission amplitude $\Psi^{\text{(1)}}_{\text{out}}(x_{1},x_{2})$, the thin lines correspond to the one photon absorption amplitude $\Psi^{\text{(2)}}_{\text{out}}(x_{1},x_{2})$, and the broken lines correspond to the two photon absorption amplitude  $\Psi^{\text{(3)}}_{\text{out}}(x_{1},x_{2})$.}
\end{figure}

In order to illustrate the dependence of the output component on the relative distance between two photons in detail, figure.~\ref{fig:figure9} shows the cross-sections at $\left|x_{1}-x_{2}\right|=c\tau$ of the output amplitude $\Psi_{\text{out}}(x_{1},x_{2})$ (solid line) and the components of the transmission process $\Psi_{\text{out}}^{\text{(1)}}(x_{1},x_{2})$ (dotted line), the one photon absorption process $\Psi^{\text{(2)}}_{\text{out}}(x_{1},x_{2})$ (thin line), and the two photon absorption process $\Psi^{\text{(3)}}_{\text{out}}(x_{1},x_{2})$ (broken line). The time differences are $\tau=0$ in (a), $\tau=0.3/\Gamma$ in (b), and $\tau=1/\Gamma$ in (c). In all figures, the component of the one photon absorption process (thin line) is always negative and significantly greater than the component of the transmission process (dotted line). Furthermore, the ratio of the peak of the one photon absorption amplitude (thin line) to the peak of the corresponding transmission amplitude does not change very much depending on the time difference $\tau$ between the two photons. On the other hand, the ratio of the peak of the two photon absorption amplitude (broken line) to the peak of the corresponding transmission amplitude (dotted line) changes significantly depending on the time difference $\tau$ between the two photons as discussed in the following.

Figure.~\ref{fig:figure9} (a) shows the cross-section of the output components at $\tau=0$. In this case, the two photon absorption amplitude (broken line) is exactly zero. The output amplitude is then given by the sum of the transmission amplitude (dotted line) and the one photon absorption amplitude (thin line). Since the one photon absorption amplitude is about $2.5$ times as large as the transmission amplitude, the total output pulse (solid line) is negative and about $1.5$ times as large as the transmission amplitude.

Figure.~\ref{fig:figure9} (b) shows the cross-sections of the output components at $\tau=0.3/\Gamma$. The one photon absorption amplitude (thin line) and the transmission amplitude (dotted line) are not much different from $\tau=0$ (figure.~\ref{fig:figure9} (a)). However, the two photon absorption amplitude (broken line) is now about half as high as the transmission amplitude (dotted line). Therefore, the output amplitude is reduced to about the same height as the transmission amplitude.

Figure.~\ref{fig:figure9} (c) shows the cross-sections of the output components at $\tau=1/\Gamma$. The two photon absorption amplitude (broken line) and the transmission amplitude (dotted line) are now significantly lower than for $\tau=0$ (figure.~\ref{fig:figure9} (a)). However, the negative one photon absorption amplitude is still about $3$ times as large as the transmission amplitude. The two photon absorption amplitude (broken line) has increased to about $1.5$ times the transmission amplitude, leaving a total negative output amplitude (solid line) of about $0.5$ times the transmission amplitude. Thus the output amplitude is reduced to very low values, even though the components themselves are still very large. The fast decrease of the total output amplitude shown in figure.~\ref{fig:figure7} (c) and \ref{fig:figure7} (d) can then be understood as a consequence of the rapid increase of the two photon absorption amplitude with $\tau$.

Interestingly, the delay times of the pulse components are not sufficiently high to cause the same change in the sign of the amplitude observed for the one photon case (figure.~\ref{fig:figure6} (a)). No part of the two-photon output wavefunction is dominated by either the transmission or the two photon absorption amplitude. This is mostly because the two photon absorption amplitude is suppressed by the non-linearity. In fact, it can be seen in figure.~\ref{fig:figure9} (c) that the nonlinearity even reduces the delay of the two photon absorption amplitude from a delay time of $0.6/\Gamma$ for $\Psi_{\text{abs}}(x_{1}) \cdot \Psi_{\text{abs}}(x_{2})$ to only about $0.45/\Gamma$ for $\Psi^{\text{(3)}}_{\text{out}}(x_{1},x_{2})$ at $\tau=1/\Gamma$. Therefore, the change of the pulse shape in the two photon pulses is actually less drastic than the change of pulse shape observed in the one photon case.

\section{\label{sec:level5}SUMMARY AND CONCLUSION}
We have studied the effect of the strong nonlinearity of a one-dimensional atom on one- and two-photon input wavepakets. The response of the atom can be characterized by transmission and absorption processes. For long pulses, the absorption components dominate in the output. However, in the two photon case, the two photon absorption component is suppressed around $x_{1}=x_{2}$ by the saturation of the two-level atom. The long two photon pulse therefore has a non-linear region of negative amplitude dominated by the one photon absorption for $\left|x_{1}-x_{2}\right| < 1.4c/\Gamma$, and a linear region of positive amplitude dominated by the two photon absorption for $\left|x_{1}-x_{2}\right| > 1.4c/\Gamma$. For a short pulse length of $T=1/\Gamma$, the two photon absorption is strongly suppressed throughout the whole output pulse. The one photon absorption process then dominates everywhere, and the amplitude of the short two-photon output pulse is completely negative. It is therefore possible to obtain a very strong non-linear effect by adjusting the pulse length $T$ to the dipole relaxation time of the atom. 

Besides the total amplitude, we have also characterized the change in delay time and pulse shape. For long pulses, the delay of the non-linear region is reduced to $1/3$ of the delay of the liner region. In the long pulse case, the interaction of the two photons thus significantly reduces the average delay time caused by the absorption and reemission processes at the atom and causes a faster reemission into the output field. In the case of a short one photon pulse, the delay between the transmission and the absorption component is so great that the pulse shape is changed significantly. In particular, the single photon output amplitude has both negative and positive regions. On the other hand, the output pulse shape for a short two photon pulse is changed much less due to the reduced delay caused by the non-linearity. It is therefore possible to obtain a negative output amplitude throughout the short output pulse. This result indicates that the atomic nonlinearity is particularly strong at a pulse length of $T=1/\Gamma$.

The analysis of the output in terms of components corresponding to different interaction processes thus allow us to understand the origin of the non-linear photon-photon interaction in great detail. These insights may be very useful for the development of novel non-linear optical devices operating at the few photon level.

\vspace{0.5cm}
\begin{flushleft}
\textit{\textbf{\large Acknowledgements}}
\end{flushleft}

This work was partly supported by the program "Research and Development on Quantum Communication Technology" of the Ministry of Public Management, Home Affairs, Posts and Telecommunications of Japan.
\bibliographystyle{plain}

\begin{thebibliography}{99}
\bibitem{nielsen0}Nielsen,~M., Chuang,~I. (2000). {\it Quantum Computation and Quantum Information.} Cambrige: Cambridege University Press.
\bibitem{holger3}Hofmann,~H., Kojima,~K., Takeuchi,~S., Sasaki,~K. (2003). Entanglement and four-wave mixing effects in the dissipation-free nonlinear interaction of two photons at a single atom. {\it Phys.~Rev.~A, 68}, 043813.
\bibitem{nondemo}Milburn,~G., Walls,~D. (1983). Quantum nondemolition measurements via quadratic coupling. {\it Phys.~Rev.~A, 28}, pp.~2065-2070; N.~Imoto, H.~A.~Haus, Y.~Yamamoto. (1985). Quantum nondemolition measurement of the photon number via the optical Kerr effect. {\it Phys.~Rev.~A, 32}, pp.~2287-2292.
\bibitem{mills}Mills,~D. (1991). {\it Nonlinear Optics (Basic Concepts).} New York: Springer-Verlag.
\bibitem{bermann}Bermann,~P. (1994). {\it Cavity Quantum Electrodynamics Supplement 2 to Advances in Atomic, Molecular, and Optical Physics.} San Diego: Academic.
%\bibitem{fredkin}G.~J.~Milburn. (1989). {\it Phys.~Rev.~Lett {\bf 62}}, pp.~2124.
\bibitem{turchette}Turchette,~Q., Hood,~C., Lange,~W., Mabuchi,~H., Kimble,~H. (1995). Measurement of Conditional Phase Shifts for Quantum Logic. {\it Phys.~Rev.~Lett., 75}, pp.~4710-4713.
\bibitem{holger2}Hofmann,~H., Kojima,~K., Takeuchi,~S., Sasaki,~K. (2003). Optimized Phase Switching using a single-atom nonlinearity { \it J.~Opt.~B: Quantum.~Semiclass.~Opt., 5}, pp.~218-221.
\bibitem{kojima}Kojima,~K., Hofmann,~H., Takeuchi,~S., Sasaki,~K. (2003). Nonlinear interaction of two photons with a one-dimensional atom: Spatiotemporal quantum coherence in the emitted field. {\it Phys.~Rev.~A, 68}, 013803.
\bibitem{holger0}Hofmann,~H., Mahler,~G. (1995). Measurement models for time-resolved spectroscopy: a comment. { \it Quantum.~Semiclass.~Opt. 7}, pp.~489-497.
\bibitem{rice}Rice,~P., Carmichael,~H. (1988). Single-Atom Cavity-Enhanced Absorption I: Photon Statistics in the Bad-Cavity Limit. {\it IEEE J.~Quantum~Electron., 24}, pp.~1351-1365.
\bibitem{turchetteb}Truchette,~Q., Thompson,~R., Kimble,~H. (1992). One-dimensional atoms. {\it Appl.~Phys.~B, 60}, pp.~S1-S10.
\end{thebibliography}

\end{document}